\documentclass{aa}  

\usepackage{graphicx}
\usepackage{microtype}
\usepackage{txfonts}
\usepackage{textcomp}
\usepackage{hyperref}
\newcommand{\micron}{\textmu{m}}
\newcommand{\sbs}[1]{\ensuremath{_\text{#1}}}  
\newcommand{\sps}[1]{\ensuremath{^\text{#1}}}  
\newcommand{\summe}{\sbs{sum}}
\newcommand{\p}{\sbs{p}}
\renewcommand{\t}{\sbs{t}}
\newcommand{\ACE}{ACE} 
\newcommand{\total}{\ensuremath{\mathrm{d}}}

\begin{document}

   \title{Collisions and drag in debris discs with eccentric parent belts}

   \author{T. Löhne\inst{1}
          \and
          A.\,V. Krivov\inst{1}
          \and
          F. Kirchschlager\inst{2}%
          \and
          J.\,A. Sende\inst{1}
          \and
          S. Wolf\inst{2}%
          }

   \institute{Astrophysikalisches Institut und Universitätssternwarte,
              Friedrich-Schiller-Universität Jena,
              Schillergässchen 2--3, 07745 Jena, Germany\\
              \email{torsten.loehne@uni-jena.de}
              \and
              Institut für Theoretische Physik und Astrophysik,
              Christian-Albrechts-Universität zu Kiel,
              Leibnizstraße 15, 24118 Kiel, Germany
             }
   \date{}

 
  \abstract
   {High-resolution images of circumstellar debris discs reveal off-centred rings that indicate past or ongoing perturbation, possibly caused by secular gravitational interaction with unseen stellar or substellar companions. The purely dynamical aspects of this departure from radial symmetry are well understood. However, the observed dust is subject to additional forces and effects, most notably collisions and drag.}
   {To complement the studies of dynamics, we therefore aim to understand how new asymmetries are created by the addition of collisional evolution and drag forces, and existing ones strengthened or overridden.}
   {We augmented our existing numerical code ``Analysis of Collisional Evolution'' (\ACE) by an azimuthal dimension, the longitude of periapse. A set of fiducial discs with global eccentricities ranging from 0 to 0.4 is evolved over giga-year timescales. Size distribution and spatial variation of dust are analysed and interpreted. The basic impact of belt eccentricity on spectral energy distributions (SEDs) and images is discussed. }
   {We find features imposed on characteristic timescales. First, radiation pressure defines size cutoffs that differ between periapse and apoapse, resulting in an asymmetric halo. The differences in size distribution make the observable asymmetry of the halo depend on wavelength. Second, collisional equilibrium prefers smaller grains on the apastron side of the parent belt, reducing the effect of pericentre glow and the overall asymmetry. Third, Poynting--Robertson drag fills the region interior to an eccentric belt such that the apastron side is more tenuous. Interpretation and prediction of the appearance in scattered light is problematic when spatial and size distribution are coupled.}
   {}


   \maketitle

\section{Introduction}
Observed dust in debris discs is produced in collisions amongst orbiting planetesimals.
Resolved images at submillimeter wavelengths, which are tracing large grains, suggest the
dust parent bodies to be arranged in narrow belts, similar to the classical Kuiper belt
in the Solar system. These narrow planetesimal belts are also evident in scattered light
images in the optical and near-infrared. The small dust grains visible at these wavelengths
are most abundant at the same locations where their parent bodies reside.

In some of the debris discs, these narrow planetesimal belts appear eccentric and
show a global offset between the belt centre and the star.
The disc in the Fomalhaut A system \citep{kalas+2005,kalas+2013}
is perhaps the most prominent example. Another example is the HD~202628 disc
\citep{krist+2012}.
Many other discs, such as HD~32297 \citep{kalas2005}, HD 61005 \citep[also known as ``The Moth'',][]{hines+2007},
and HD~15115 \citep[``The Blue Needle'',][]{kalas+2007a},
exhibit a global asymmetry between the two wings.
It is possible that these asymmetry also derives from the offsets in the underlying belts,
which may not be seen because of the edge-on orientation of the discs and/or
insufficient spatial resolution of the submillimeter facilities.
Belt offsets can naturally be explained by as yet undiscovered planets
in eccentric orbits interior to the belts \citep{lee+chiang2016,esposito+2016}
or substellar companions exterior to them \citep{thebault+2010,thebault2012,nesvold+2016}.
Alternatively, the wing asymmetries may also be caused by 
recent giant collisions \citep[e.\,g.,][]{kral+2015,olofsson+2016} or
displacement of the dust by the surrounding interstellar gas \citep{debes+2009}
or dust \citep{artymowicz+clampin1997}.

Interpretation of asymmetries in discs in terms of potential perturbers
requires models to predict how exactly such planets would shape the distribution
of the disc material. Whereas the influence of planets on the parent belts is easily
understood with the Laplace--Lagrange secular perturbation theory, the task gets more complicated
for small dust grains. These dominate the cross-section and thus also 
the observable appearance of extrasolar debris discs.
However, these are no direct tracers of the underlying distribution of parent bodies
from that they are produced, because they are subject to an additional array
of forces and effects, including collisional production and removal,
radiation pressure, and drag forces \citep[e.\,g.,][]{wyatt+1999}.


Previous work 
\citep[among others]{stark+kuchner2008,stark+kuchner2009,kuchner+stark2010,thebault+2012,%
vitense+2012,kral+2013,nesvold+2013,thebault+2014,vitense+2014,kral+2015,nesvold+kuchner2015a,%
nesvold+kuchner2015b,lee+chiang2016,esposito+2016}
extended purely gravitational models of planet-disc interactions
by including these effects and forces acting on dust grains.
In this paper, we tackle the problem with a novel approach that is based on modelling of the
evolution of the phase space distributions of the material, rather than N-body integrations as was done previously.
We show that a combination of
grain-grain collisions and Poynting-Robertson drag
with the gravitational perturbations by massive bodies in the system
creates inseparable size-spatial distributions of solids.
These are used to explore the distributions of dust grain sizes in different disc locations
and expected observable signatures in discs.

We start with a discussion of secular perturbations by planets as a potential cause for
the narrow eccentric belts in Section~\ref{sec:Origins}.
In Section~\ref{sec:Model}
we introduce a new version of our collisional code \ACE\ that can now treat
azimuthal asymmetry. Sections~\ref{sec:Distributions} and
\ref{sec:Observables} present the outcomes of simulations of fiducial
discs in terms of dust distributions and observables, respectively.
We summarise our findings in Section~\ref{sec:Summary}.

\section{On possible origins of the offsets}\label{sec:Origins}

The presence of eccentric belts is evident from observed images, but the origin of these offsets is yet unclear.
Mechanisms other than a planet in eccentric orbit in the disc's cavity are conceivable. For instance, \citet{shannon+2014} propose that the eccentricity of the belt around Fomalhaut A was set by dynamical interactions with the other two companions of this triple system, Fomalhaut~B and~C. Yet the planetary scenario is considered the most generic, and we now address it in more detail.

\subsection{Secular perturbations}\label{sec:Origins:Perturbations}
\begin{figure}
  \centering
  \includegraphics[width=0.45\hsize]{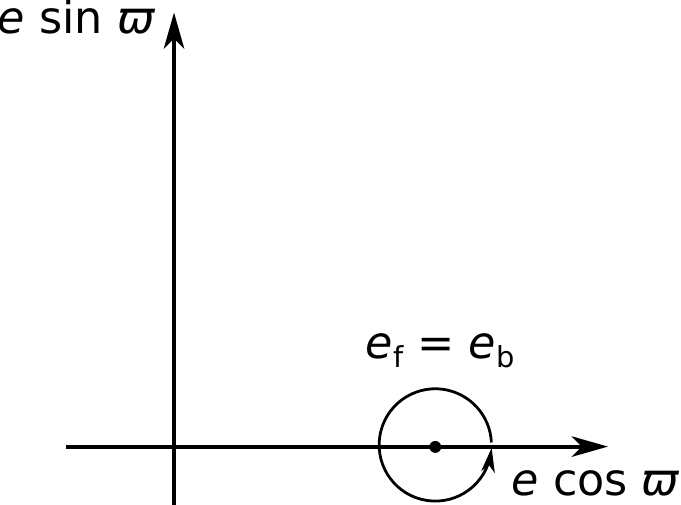}~\includegraphics[width=0.45\hsize]{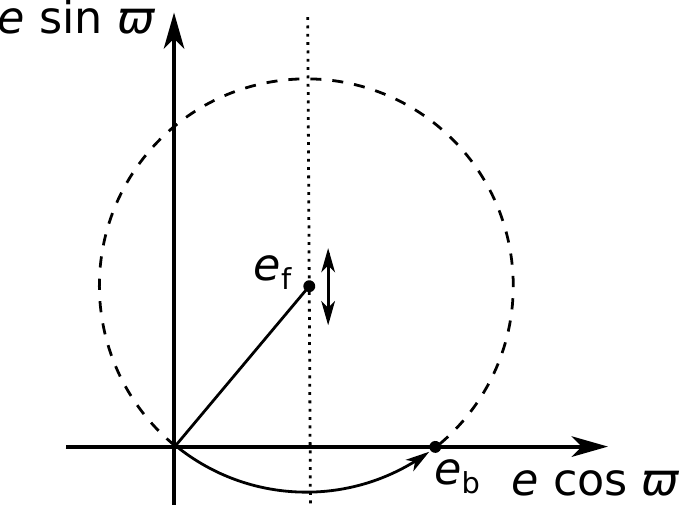}
  \caption{Possible scenarios for the origin of eccentric narrow belts through secular perturbation: \emph{(left)} equilibrium precession of the complex eccentricities around a forced eccentricity $e\sbs{f}$ close to the observed average belt eccentricity $e\sbs{b}$, \emph{(right)} ongoing precession around an unknown $e\sbs{f}$ from zero to a currently observed value $e\sbs{b}$.}
  \label{fig:ForcedEccentricity}
\end{figure}

\begin{figure}
  \centering
  \includegraphics{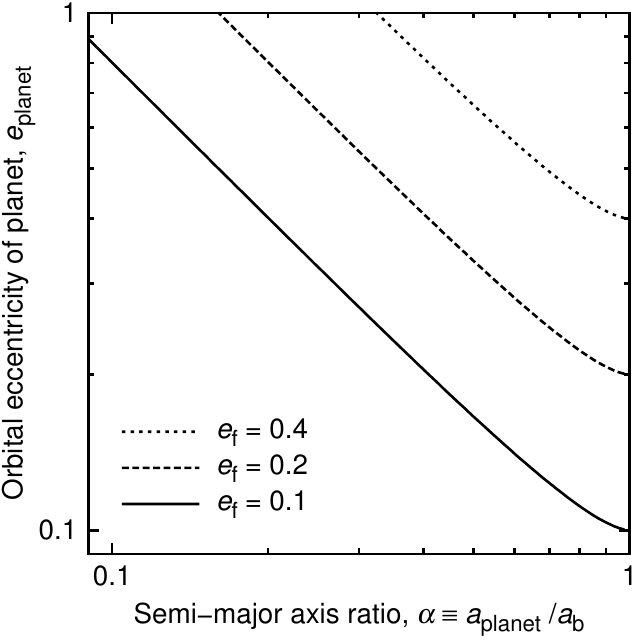}
  \caption{Planetary orbital eccentricity that is necessary to produce a given forced eccentricity $e\sbs{f}$ as a function of the ratio of semi-major axes between planet and perturbed belt.}
  \label{fig:LaplaceEccRatio}
\end{figure}

Where both short-period perturbations and resonances are unimportant, secular Laplace--Lagrange theory provides the means to compute the perturbing influence of planets or substellar companions \citep[see, e.\,g.,][]{murray+dermott2000}. Differences between mean anomalies of perturber and perturbee are assumed random in that approximation. While no energy is exchanged, orbital eccentricities and orientations of the orbits change. In a space spanned radially by eccentricity $e$ and azimuthally by longitude of periapse $\varpi$ (Fig.~\ref{fig:ForcedEccentricity}), secular perturbation makes the eccentricity vector $(e\cos\varpi, e\sin\varpi)$ precess uniformly along a circle with a radius called proper eccentricity, $e\sbs{p}$, centred around a forced eccentricity $e\sbs{f}$
\citep{hirayama1918}. The closer the combination of $e\sbs{p}$ and $e\sbs{f}$ gets to unity, the higher are the deviations from perfect circles \citep[e.\,g.,][]{beust+2014}. The forced eccentricity vector is aligned parallel to the planet's eccentricity vector, its absolute value being \citep{murray+dermott2000}:
\begin{equation}
  e\sbs{f} = e\sbs{planet} \frac{b_{3/2}^{(2)}(\alpha)}{b_{3/2}^{(1)}(\alpha)} = \left[\frac{5}{4} \alpha + o(\alpha)\right] e\sbs{planet},
\end{equation} 
where $e\sbs{planet}$ is the absolute eccentricity of the planet's orbit.
The $b$'s are Laplace coefficients, which only depend on the ratio of semi-major axes of the (interior) planet and a perturbed belt object, $\alpha \equiv{a\sbs{planet}}/a\sbs{b} < 1$. A given forced eccentricity can be caused by a nearby planet of the same eccentricity or a closer-in planet of higher eccentricity. Figure~\ref{fig:LaplaceEccRatio} depicts the corresponding planetary orbital eccentricity as a function of its relative distance to the perturbed belt. Although better approximations exist for perturber eccentricities $e\sbs{planet}>0.2$ \citep[see, e.\,g.,][and references therein]{mustill+wyatt2009}, we use Laplace--Lagrange theory for the broad analysis in this section.

While the perturber's mass does not influence $e\sbs{f}$ (as long as $M\sbs{planet} \ll M_*$), it determines the timescale on which this precession occurs. Based on the (approximate) angular precession frequency for a single perturber \citep{murray+dermott2000},
\begin{equation}
  A = \sqrt{\frac{GM_*(1 - \beta)}{a\sbs{b}^3}} \frac{M\sbs{planet}}{4M_*(1 - \beta)}\alpha b_{3/2}^{(1)}(\alpha),
\end{equation}
the time for a full precession cycle can be estimated from
\begin{equation}
  T\sbs{prec}\sps{full} \equiv \frac{2\pi}{A} = \frac{M_*(1 - \beta)}{M\sbs{planet}}\frac{4 P\sbs{b}}{\alpha b_{3/2}^{(1)}(\alpha)},\label{eq:PrecessionCycle}
\end{equation}
where $P\sbs{b} = 2\pi \sqrt{a\sbs{b}^3/[GM_*(1 - \beta)]}$ is the orbital period in the belt. The additional factor $1 - \beta$ with $\beta \equiv F\sbs{pr}/F\sbs{grav}$, the ratio of the forces due to radiation pressure and gravitational pull, accounts for the reduction of effective stellar mass because of radiation pressure.

An important criterion that the dynamical evolution must fulfil is to allow for narrow belts.
We consider two principle possibilities (Fig.~\ref{fig:ForcedEccentricity}), a simplification of the scheme of four classes discussed by \citet{thilliez+maddison2015}. One is that planetesimals in the belt have low proper eccentricities, $e\sbs{p}$.
In equilibrium, over timescales longer than the precession period, the width of the belt is then set by these proper eccentricities. If they are low, the belt will remain narrow at all times (Fig.~\ref{fig:ForcedEccentricity} left). However, this raises the question of how the orbits of the parent bodies came close to their forced value in the $e$-$\varpi$ plane. It would be more natural to assume a second possibility, in which planetesimals were born in nearly circular orbits, before the planetary perturber emerged in the protoplanetary disc. If so, we must require the precession circle to cross the centre (i.\,e., the point $e=0$) and the currently observed average belt eccentricity $e = e\sbs{b}$. The minimum time for the belt eccentricity to precess to $e\sbs{b}$ is a fraction
\begin{equation}
  T\sbs{prec} \approx  T\sbs{prec}\sps{full} \frac{e\sbs{b}}{2\pi e\sbs{f}}
\end{equation}
of a full cycle (see Fig.~\ref{fig:ForcedEccentricity} right), potentially increased by $NT\sbs{prec}\sps{full}$ if $N$ precession cycles have already passed. However, precession will also smear out the orbits around the forced eccentricity in a wide circle if the belt had no offset prior to the perturbation. The resulting spread in $e$--$\varpi$ would be twice as large as $e\sbs{f}$ itself, leading to wide instead of narrow belts \citep{beust+2014}. Thus, for this scenario to produce a narrow belt, the differential precession timescales must be longer than the time since perturbation started, in order to prevent this smearing out. We can define such a differential timescale as
\begin{equation}
  \Delta T\sbs{prec} \equiv \frac{\Delta e\sbs{b}}{\Delta A},\label{eq:DeltaTprec}
\end{equation}
where $\Delta e\sbs{b}$ is the spread in eccentricities and $\Delta A$ denotes the spread in linear change rates in eccentricity space:
\begin{equation}
  \Delta A = \frac{\total (A e\sbs{f})}{\total \alpha} \frac{\total \alpha}{\total a\sbs{b}} \Delta a\sbs{b}
           = A e\sbs{f} \frac{\Delta a\sbs{b}}{a\sbs{b}} \left[\frac{5}{2} + \frac{\alpha}{b_{3/2}^{(2)}} \frac{\total b_{3/2}^{(2)}}{\total\alpha}\right].\label{eq:DeltaAef}
\end{equation}
Inserting (\ref{eq:DeltaAef}) into (\ref{eq:DeltaTprec}) and expanding the term in brackets in a series in $\alpha$, we obtain
\begin{equation}
  \Delta T\sbs{prec} = \frac{\Delta e\sbs{b}}{A e\sbs{f}} \frac{a\sbs{b}}{\Delta a\sbs{b}} \left[\frac{9}{2} + \frac{7}{2}\alpha^2 + \frac{119}{32}\alpha^4 + o(\alpha^4)\right]^{-1},
\end{equation}
or with equation~(\ref{eq:PrecessionCycle}),
\begin{equation}
  \Delta T\sbs{prec} \approx T\sbs{prec}\sps{full} \frac{\Delta e\sbs{b}}{2\pi e\sbs{f}} \frac{a\sbs{b}}{\Delta a\sbs{b}} \frac{1 - \alpha^2}{4}
\end{equation}
because the coefficients in the series approach a value of $4$. The spread in eccentricities ($\Delta e\sbs{b}$) remains lower than the average eccentricity ($e\sbs{b}$) and the belt remains narrow as long as $\Delta T\sbs{prec} > T\sbs{prec}$. We find
\begin{equation}
  \frac{\Delta e\sbs{b} = e\sbs{b}}{2\pi e\sbs{f}} \frac{a\sbs{b}}{\Delta a\sbs{b}} \frac{1 - \alpha^2}{4} > \frac{e\sbs{b}}{2\pi e\sbs{f}} + N
\end{equation}
and
\begin{equation}
  \frac{\Delta a\sbs{b}}{a\sbs{b}} < \frac{1 - \alpha^2}{4} \left(1 + \frac{2\pi N e\sbs{f}}{e\sbs{b}}\right)^{-1}.
\end{equation}
This constraint on the belt width is mild for $N = 0$ and small values of $\alpha$, but strong for $\alpha$ close to $1$ and $N > 0$. That is, it is more likely that an observed narrow belt is still in its first precession cycle. Assume, for example, that the belt is distant from the planet, $\alpha \ll 1$, and that the observed belt eccentricity is approximately the same as the forced eccentricity. We find $\Delta a\sbs{b}/a\sbs{b} < 1/4$ for $N = 0$ and $\Delta a\sbs{b}/a\sbs{b} \lesssim 1/29$ for $N = 1$. While the first belt can be broad, the second belt needs to be narrow.



\subsection{Constraints on perturbing planets}\label{sec:Origins:Planets}

Not excluding either of the two possibilities described in Section~\ref{sec:Origins:Perturbations}, we now briefly discuss what both would mean for the unseen perturbing planet. In the modelling described in the rest of the paper, we use the mean eccentricity of the parent belt, $e\sbs{b}$, as a key paremeter. However, its interpretation in these two cases is different. In the low-$e\sbs{p}$ scenario, $e\sbs{b}$ is equal to the forced eccentricity $e\sbs{f}$ (Fig.~\ref{fig:ForcedEccentricity} left). In that case, the planet orbit's apsidal line is aligned with the major axis of the belt. In the slow-precession scenario, $e\sbs{b}$ is {\em not} equal to $e\sbs{f}$. Instead, it represents the instantaneous value of the complex eccentricity $e$ (Fig.~\ref{fig:ForcedEccentricity} right). In that case, the planetary orbit will be misaligned with the major axis of the belt \citep{beust+2014}.

\section{Collisional model}\label{sec:Model}
The number of particles in debris discs is orders of magnitude beyond the scope of pure $N$-body simulations. Hence, statistical representations for particle distributions and/or collisions are used. Collision rates and outcomes are calculated for whole groups of similar particles, called super-particles or bins or tracers or streamlines. A major difficulty common to all approaches is the sampling: the number of groups needs to be high enough to properly represent the modelled distribution, and low enough to be computationally tractable. There are two main approaches to this grouping: (A) time-resolved and (B) orbit-averaged.

In approach A, particles in close spatial proximity and with similar velocity vectors are grouped into so-called super-particles \citep{grigorieva+2007a}, which can be viewed as more or less coherent clouds of particles that move in parallel. When two clouds collide, collision rates among individual particles are calculated based on the local particle-in-a-box principle. The collision cross-section -- or the volume of interaction -- of the super-particles can either be defined as a co-moving sphere \citep[e.\,g.,][]{grigorieva+2007a} or a (revolving) grid element in polar coordinates \citep[e.\,g.][]{kral+2013}. Smaller interaction volumes reduce the rate of collisions per super-particle, but increase the rate of individual collisions per super-collision. Smaller super-particles allow for higher spatial (and temporal) resolution, but require more super-particles to reduce noise artefacts. The biggest advantage of this approach is the ability to model short-term effects, such as collisional avalanches \citep{grigorieva+2007a}, major break-ups of planetesimals \citep[with LIDT-DD]{kral+2015} or close stellar flybys \citep[with SMACK]{nesvold+2017}. Due to the underlying $N$-body integration, additional forces are easily implemented in these codes, including radiation pressure, drag-induced spiralling, resonant capture, and scattering. ``Collisional grooming'', the algorithm presented by \citet{stark+kuchner2009}, can be considered a variant in which collisions do not create new dust. Once released at a (constant) production rate, grains stream along their trajectories, while their number densities are gradually reduced in collisions as their trajectories cross others (or themselves). The model settles towards an equilibrium. LIPAD \citep{levison+2012} is a further example for approach A.

In approach B, which our code \ACE\ follows, particles are grouped according to their orbits. Instead of local clouds, each group populates a given ellipse (or hyperbola), with particle density uniform accross mean anomalies, i.\,e. uniform in time. Discrete orbits are fixed throughout the simulations, parameterised either by orbital elements (\ACE), or again, by location and velocity vector \citep{thebault+2003}. This orbit-averaging makes the models meaningful only on timescales longer than the orbital period. It excludes application to very dense discs with short collision timescales and to the short-term perturbations that can be analyzed with approach A. On the other hand, the phase space of orbital elements eases long-term computations with time steps of millions of years, i.\,e. orders of magnitude longer than orbital periods.


\subsection{Phase space and master equation}
While previous versions of \ACE\ could only treat axisymmetric debris discs \citep[e.\,g.,][]{krivov+2006,reidemeister+2011,loehne+2012,krivov+2013}, the azimuthal distribution is now allowed to be non-uniform. The phase space spans an additional dimension that covers the orientations of object orbits, parameterized by longitude of periapsis $\varpi \equiv \Omega + \omega$, the sum of the longitude of the ascending node ($\Omega$) and the argument of periapsis ($\omega$). The vertical dimension is still averaged over \citep[cf.][]{krivov+2006}.

In total, the discretized distribution of material is represented by bins with four dimensions: (1) object masses $m$, (2) orbital pericentres $q$, (3) eccentricities $e$, and (4) longitudes of periapse $\varpi$. Collisions among pair-wise combinations of bins are possible at up to two distinct points defined by the colliders' $q$s, $e$s, and $\varpi$s. Collision velocities, rates, and outcomes follow directly \citep{krivov+2006}.

The discretized master equation to be integrated over time reads
\begin{equation}
  \dot{n}_i = \sum_{jk}\limits G_{ijk} n_j n_k - \sum_j\limits L_{ij} n_i n_j + \sum_j\limits T_{ij} n_j,
\end{equation}
where $n_i$ is the number (or mass) of objects in the bin specified by multi-index $i \equiv (i_m, i_q, i_e, i_\varpi)$.
Coefficients $G_{ijk}$ denote the gain, the specific rates at which objects of type $i$ are formed in collisions among objects of types $j$ and $k$;
$L_{ij}$ denote the loss, the specific rate at which objects of type $i$ are removed in collisions with $j$. \ACE\ models drag forces by advection from grid cell to grid cell \citep[since][]{reidemeister+2011}. The coefficients $T_{ij}$ hence denote the transport to cell $i$ from (neighbouring) cell $j$. For example, Poynting--Robertson drag reduces $q$ in bin $j$ at a rate $\dot{q}_j$. Given a bin width $\Delta q_j$, the contents of bin $j$ are moved towards $i$ at a rate $T_{ij} n_j = n_j \dot{q}_j / \Delta q_j$.

With collision physics depending on the orbits' orientations only
through the difference $\varpi_i - \varpi_j$, the dependence of $G_{ijk}$ and $L_{ij}$ on $i_\varpi$, $j_\varpi$, and $k_\varpi$ is only through pair-wise differences $i_\varpi - j_\varpi$, etc.
The relation $G_{ijk} = G_{i'j'k'}$ for $i' = (i_m, i_q, i_e, i_\varpi + o)$, $j' = (j_m, j_q, j_e, j_\varpi + o)$, and $k' = (k_m, k_q, k_e, k_\varpi + o)$ with $o \in N$ can be employed to speed up the calculations; instead of looping colliders over $j_\varpi$ and $k_\varpi$, $k_\varpi - j_\varpi$ and $j_\varpi$ are used, making the innermost loop over $j_\varpi$ trivial.

\subsection{Collision outcomes}
Depending on the colliders' masses and the impact energy, we consider three main outcomes. First, disruption and dispersal occurs if the energy suffices to overcome both the material strength and the combined gravitational potential of the colliders. For that threshold specific energy we follow \citet{loehne+2012} and assume the size dependence described by \citet{benz+asphaug1999} together with the modification from \citet{stewart+leinhardt2009}:
\begin{eqnarray}
Q_{ij}^* &=& \left[5\times 10^2~\text{J/kg}\,\left(\frac{s_{ij}}{1\,\text{m}}\right)^{-0.37}\right.\nonumber\\
                &+& \left.5\times 10^2~\text{J/kg}\,\left(\frac{s_{ij}}{1~\text{km}}\right)^{1.38}\right] \left(\frac{v\sbs{imp}}{3~\text{km/s}}\right)^{0.5}\nonumber\\
                &+& \frac{3G(m_i + m_j)}{5s_{ij}}\label{eq:Q},
\end{eqnarray}
where the two terms in brackets on the right-hand-side represent (1) shock disruption in strength regime and (2) in gravity regime, scaled by impact velocity $v\sbs{imp}$. By $s_{ij} \equiv (s_i^3 + s_j^3)^{1/3}$ we denote the equivalent radius of a sphere with the combined volume of the colliders. The last line in equation~(\ref{eq:Q}) approximates the specific energy required to overcome self-gravity. It is important only for radii $s \gtrsim 30$~km and has not been taken into account in previous \ACE\ versions.

Second, below that threshold, we call collisions cratering if the target retains at least half of its original mass, but half the impact energy is enough to disrupt the projectile. Gravitational accretion also falls into this category. Third, if both colliders stay intact, they are assumed to separate again, unless impact velocities are below 10~m/s. In all three cases, a cloud of smaller fragments is produced in addition to the colliders' remnants. In the model, the total mass in escaping fragments is proportional to impact energy:
\begin{equation}
  m\sbs{frag} = \frac{Q}{Q_*} \left(\frac{m_i}{2} + m_j \right).
\end{equation}
The fragment mass distribution is assumed to follow a power law with exponent $\eta = -11/6$ up to a limiting mass, above which the power-law distribution would accumulate to exactly one further particle. The mass of the largest fragment is thus given by
\begin{equation}
  m\sbs{lf} \equiv  \frac{2.0 + \eta}{- \eta - 1.0} m\sbs{frag} = \frac{1}{5} m\sbs{frag}.
\end{equation}
where $m\sbs{frag}$ is the total mass in escaping fragments. The same prescription was used in previous \ACE\ versions.

The remnants from erosive collisions are treated somewhat different from new fragments. If their new combination of mass and orbit still has them in the same bin, the total mass in that bin is reduced by how much is transformed into fragments. Only if the remnants move to a different bin, is their mass added to the loss of one bin and to the gain of another. However, in what follows we will simply refer to both fragments and remnants as fragments.

\subsection{Fragment orbits}\label{sec:Model:FragOrb}
Momentum conservation requires that the cloud of fragments produced in a collision
has the same centre of mass as the original colliders. Neglecting relative velocities of the fragments, i.\,e. assuming full energy dissipation, all share a common initial velocity.
As soon as the cloud becomes optically thin, radiation (and wind) pressure segregate the fragments according to their size-dependent $\beta$ ratios.
%
In the spirit of \citet{krivov+2006} the following orbital semi-major axis $a$, semilatus rectum $p$,
and eccentricity $e$ can be derived for a fragment produced in
a collision between a target (subscript t) and a projectile (subscript p) at a distance $r$:
\begin{eqnarray}
  \frac{r}{a} &=& 2 - \frac{m\p'^2}{m\summe'^2}\cdot
                      \left[2 - \frac{r}{a\p}\right]
                    - \frac{m\t'^2}{m\summe'^2}\cdot
                      \left[2 - \frac{r}{a\t}\right] \nonumber\\
            & &   - 2\frac{m\p' m\t'}{m\summe'^2}
                    \left[\frac{1}{r}\sqrt{p\p p\t} \right.\nonumber\\
   &&    \left.\pm \sqrt{\left(2 - \frac{r}{a\p} - \frac{p\p}{r}\right)
                     \left(2 - \frac{r}{a\t} - \frac{p\t}{r}\right)}\right]
       \label{eqResult_a},\\
  p &=& \frac{m\p'^2}{m\summe'^2}p\p + \frac{m\t'^2}{m\summe'^2}p\t + 2\frac{m\p'm\t'}{m\summe'^2}\sqrt{p\p p\t},   
        \nonumber\label{eqResult_p}\\
  e &=& \mathrm{sgn}(1 - \beta) \sqrt{1 - \frac{p}{a}}\label{eqResult_e},
\end{eqnarray}
where $m\t' \equiv m\t \sqrt{1 - \beta\t}$ and $m\p' \equiv m\p \sqrt{1 - \beta\p}$ denote the effective masses of target and projectile, and $m\summe' \equiv (m\t + m\p)\sqrt{1 - \beta}$ the sum of the collider masses, weighed by the $\beta$ ratio of the fragment.
The signs of $e$ and $p$ are equal to that of $1-\beta$, corresponding to anomalous
hyperbolae with $e < 0$ for $\beta > 1$.

For grains that are released from a macroscopic body with $\beta_t = 0$ on impact of a projectile with $m\p \ll m\t$, equation~(\ref{eqResult_a}) is reduced to
\begin{equation}
  a = a\t \frac{1 - \beta}{1 - 2\beta a\t/r}.
\end{equation}
The blowout limit is reached where $a \rightarrow \infty$, corresponding to \citep{kresak1976}
\begin{equation}
  \beta\sbs{lim} = \frac{r}{2a\t}.\label{eq:blowout}
\end{equation}
That limit varies between $(1+e)/2$ and $(1-e)/2$ for a grain release at
apastron ($r = Q = a(1 + e)$) and periastron ($r = q = a(1 - e)$), respectively. When launched from circular orbits, grains become unbound for $\beta \geq 1/2$.

If the material distribution is non-axisymmetric, the relative orientations of fragment orbits with respect to those of the initial colliders
need to be considered in the modelling.
The true anomaly $\theta$ of a freshly released fragment can be determined from
\begin{equation}
  e \sin\theta = \frac{m\t'}{m\summe'}\sqrt{\frac{p}{p\t}} e\t \sin\theta\t + \frac{m\p'}{m\summe'}\sqrt{\frac{p}{p\p}} e\p \sin\theta\p .
\end{equation}
and
\begin{equation}
  e \cos\theta = \frac{p}{r} - 1.
\end{equation}
At the mutual crossing points of two orbits, the difference of true anomalies equals the (negative) difference between their longitudes of periapse:
\begin{equation}
  \theta' - \theta = \varpi - \varpi'.
\end{equation}
For collisions between projectiles and targets, the true anomalies of the latter at these points
are given by \citep[cf.][eq. (8)]{krivov+2006}
\begin{equation}
  \theta\t = \arcsin \frac{A}{\sqrt{C}} \pm \arccos \frac{B}{\sqrt{C}}
\end{equation}
or
\begin{equation}
  \sin\theta\sbs{t} = \frac{AB \pm D\sqrt{C - B^2}}{C},\quad\cos\theta\sbs{t} = \frac{BD \mp A\sqrt{C - B^2}}{C},
\end{equation}
where
\begin{eqnarray}
  A &\equiv& e\p p\t\sin(\varpi\p - \varpi\t),\\
  B &\equiv& p\p - p\t,\\
  C &\equiv& e\t^2 p\t^2 + e\p^2 p\p^2 - 2 p\p p\t e\p e\t \cos(\varpi\p - \varpi\t),\\
  D &\equiv& \sqrt{C - A^2} = e\p p\t\cos(\varpi\p - \varpi\t) - e\sbs{t}p\p.
 \end{eqnarray}

\subsection{Model and grid parameters}\label{sec:Model:Setup}
The aim of our study is to identify the characteristic influence of collisions and drag on perturbed discs. In this first paper, we refrain from covering the complexity of the wide parameter space spanned by observed discs. Instead we will present and discuss results for a small set of fiducial, typical debris systems. For the central star, we choose an A3\,V main-sequence star, with mass, luminosity, and effective temperature adopted from the values reported for Fomalhaut: $M = 1.92~M\sbs{Sun}$, $L=16.6~L\sbs{Sun}$, $T=8590$~K \citep{mamajek2012}. The stellar photospheric emission is modelled with the nearest point in the PHOENIX/NextGen grid of models \citep{hauschildt+1999}. We model the material with a homogeneous mix \citep{bruggeman1935} of Astrosilicate \citep{draine2003a} and water ice \citep{li+greenberg1998} in equal volume fractions, with a bulk density of 2.35~g~cm$^{-3}$. The combination of this material with the high luminosity and radiation pressure of the early-type star sets a clear blowout limit at grain sizes of a few microns.

Asymmetries are most easily identified where narrow belts are resolved, which correspond to narrow volumes in the space of orbital elements.
The relative radial width $\Delta Q\sbs{b}/Q\sbs{b}$ at the apastron $Q\sbs{b} = a\sbs{b}(1+e\sbs{b})$ of a belt can be estimated from
\begin{equation}
  \frac{\Delta Q\sbs{b}}{Q\sbs{b}} = \sqrt{\left(\frac{\Delta a\sbs{b}}{a\sbs{b}}\right)^2 + \left(\frac{\Delta e\sbs{b}}{1 + e\sbs{b}}\right)^2}
\end{equation}
if $a$ and $e$ vary independently. A given radial HWHM of, for instance, $\delta r/r = 10$\,\% does not only limit $\Delta a/a \leq 10$\,\%, but also $\Delta e\sbs{b} = e\sbs{p} \leq 0.1$. For the parent bodies in our main simulation runs, we therefore assume initial distributions that are confined to circular regions in the $(e\cos\varpi, e\sin\varpi)$ plane, centred on $(e\sbs{b}, 0)$, with radii $e\sbs{p} = 0.1$. The values used for $e\sbs{b}$ in the individual runs range from $0.0$ to $0.4$.

\begin{figure}
  \centering
  \includegraphics[width=\hsize]{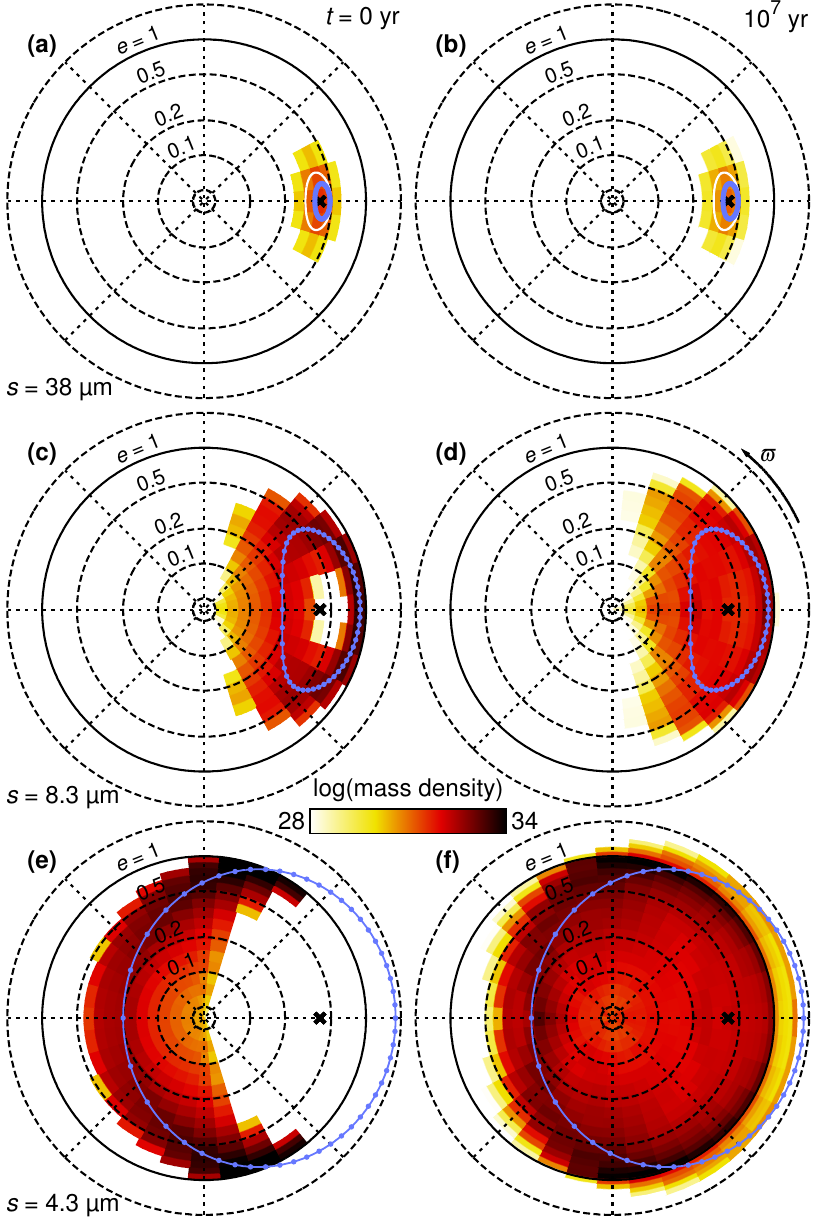}
  \caption{Phase-space distribution of grains of three different sizes for a disc with a belt eccentricity $e\sbs{b} = 0.4$:
          \emph{(top)} $s = 38$~\micron\ ($\beta = 0.06$), \emph{(middle)} $s = 8.3$~\micron\ ($\beta = 0.26$), \emph{(bottom)} $s = 4.3$~\micron\ ($\beta = 0.5$).
          Radial coordinate is logarithm of eccentricity, while the longitude of periapse $\varpi$ is plotted azimuthally.
          Color scale indicates mass per unit mass, eccentricity, and angle.
          The two columns of panels represent \emph{(left)} the initial stage at $t \approx 0$~yr and \emph{(right)} an intermediate stage at $t = 10^7$~yr.
          Black crosses mark $e\sbs{b}$.
          Light blue dots mark orbits of fragments that are launched in equidistant steps of true anomaly from a parent belt
          with zero relative velocities; the spread of blue dots around black crosses is the result of radiation pressure,
          see Sect.~\ref{sec:Model:FragOrb}.
          White lines in the top panels roughly enclose the initial distributions in the parent belt (i.\,e. the assumed
          spread $\Delta e\sbs{b}$ of parent body eccentricities around $e\sbs{b}$.
  }
  \label{fig:m_of_e_and_w}
\end{figure}

The eccentricity grid that we use spans $[0.015, 1.5]$ and is logarithmically spaced for $e\lesssim 0.4$ and for $e > 1$, with factors $\Delta e/e \approx 0.25$. In order to preserve accuracy for barely bound grains, the step size is additionally limited to $\Delta e \leq 0.1$ for $e < 1$, resulting in nearly linear steps in the range $0.4 \lesssim e < 1$. The transitions between linear and logarithmic regimes are smooth. In total, we model the number of bins per unit logarithm of eccentricity with
\begin{equation}
  \frac{\total i}{\total \ln e} = \left\{\left(\ln 1.25\right)^{-3} + \left[\left(0.1/e\right)^3 + \left(0.1/1.5\right)^3\right]^{-1} \right\}^{1/3}.
\end{equation}
The resulting number of eccentricity bins is 26. The grid of orbit orientations has 32 linear steps in $\varpi$, covering $[0, 2\pi)$. The resolution elements $\Delta e \times e\Delta\varpi$ thus measure $0.25 e \times 0.2 e$ for $e \lesssim 0.4$ and $0.1 \times 0.2 e$ for $0.4 \lesssim e < 1$. Figure~\ref{fig:m_of_e_and_w} illustrates the sub-grid of $e$ and $\varpi$ in a polar plot.

In the radial direction, a grid of logarithmically spaced periastron distances $q \equiv a(1-e)$ is used. A number of 85 bins spans a distance range from 30 au to 600 au, corresponding to relative bin widths of 0.036. In combination with an eccentricity spread $e\sbs{p} = 0.1$, this width ensures that orbits from neighbouring $q$ annuli cross. In all runs, the $q$-and-$e$ sub-grid is filled initially such that a range of semi-major axes from 95 to 110 au is covered.


Finally, masses are binned logarithmically from a minimum grain radius of $0.26$~\micron\ to a maximum of $49$~km with factors of 12 in mass (or 2.3 in radius) between adjacent bins. For grains with radii $s \lesssim 30$~µm (corresponding to $\beta \gtrsim 0.1$), where radiation pressure is important, the spacing is refined to factors of $12^{1/4}$ in mass (or 1.23 in radius), with a smooth transition between these regimes. A number of 48 mass bins result. The total number of bins in the grid that are actually filled increases over time and then saturates at about $10^6$ in the runs presented here.

The initial size distribution is assumed to follow a power law $n(s) \propto s^\nu T\sbs{orbit}/T_0$ with $\nu = -3.66$, normalized to a total mass of 2 earth masses. We follow \citet{strubbe+chiang2006} and \citet{lee+chiang2016} in scaling the initial abundances of grains with their orbital timescales $T\sbs{orbit}$ with respect to that of large grains ($\beta = 0$), $T_0$. This scaling is meant to account for the increased lifetimes of smaller grains as a result of their spending most of the time close to their apocentres, i.\,e. far from the star. We are free to choose this initial setup to ease comparison with other work. That choice does not influence the dust distribution towards which the subsequent collisional evolution will quickly converge.


\section{Resulting size and radial distributions}\label{sec:Distributions}
\begin{figure}
  \centering
  \includegraphics{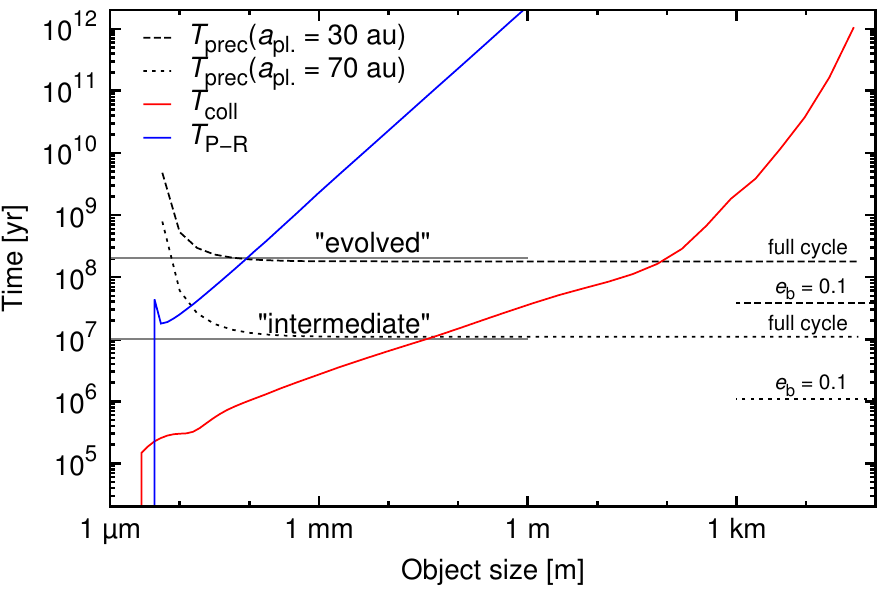}
  \caption{Average lifetimes against (solid red) collisional disruption and (solid blue) P--R drag $e$-folding times ($a/\dot a$ for $q = a(1 - e) = 100$~au and $e = \beta/(1 - \beta)$; see Sect.~\ref{sec:Distributions:Drag}) of objects in our \ACE\ run with a circular parent belt ($e\sbs{b} = 0.0$). Dashed and dotted black curves show size-dependent time scales for full precession cycles caused by planets with $M\sbs{pl} = 5\times 10^{-5} M_* = 32 M_\oplus$ at 30 and 70~au, respectively. The two short lines of the same styles illustrate the corresponding times to precess just to $e\sbs{b} = 0.1$.}
  \label{fig:Lifetimes}
\end{figure}

In this section, we will identify the impact that radiation pressure (Sect.~\ref{sec:Distributions:Dynamics}), collisions (Sect.~\ref{sec:Distributions:Collisions}), and drag forces (Sect.~\ref{sec:Distributions:Drag}) have on the spatial and size distribution of dust. Figure~\ref{fig:Lifetimes} shows the different timescales for these effects. Initially, grains populate the elliptic orbits induced by radiation pressure on short, orbital timescales. At an ``intermediate'' stage ($t = 10^7$ yr), enough time has passed to bring grains with sizes $s \lesssim 10$~cm to collisional equilibrium. Later, at an ``evolved'' stage ($t = 2\times 10^8$ yr), grains with sizes $s \lesssim 100$~\micron\ are in P--R drag equilibrium, which means that the system is older than the time these grains need to spiral from the belt to the star. In what follows, we use these three stages to illustrate the different effects. A more detailed comparison of time scales can be found in Section~\ref{sec:Distributions:Timescales}.

\subsection{Dynamical consequences}\label{sec:Distributions:Dynamics}
\begin{figure}
  \centering
  \includegraphics{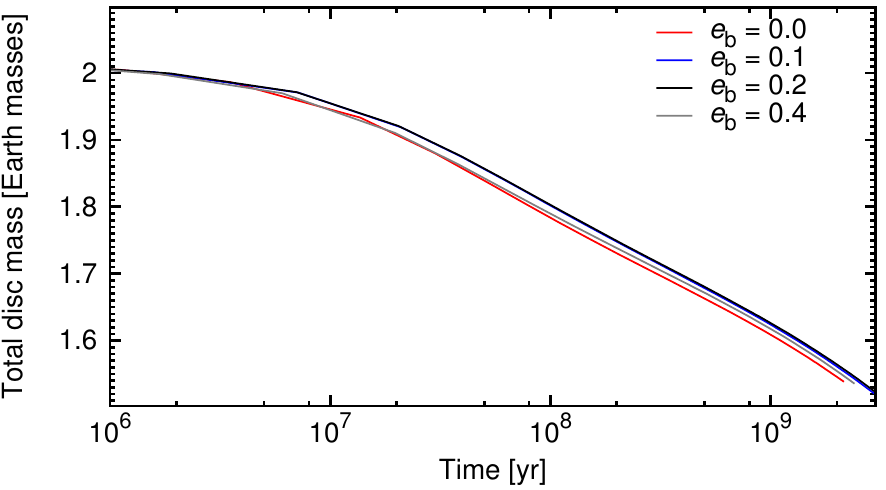}
  \caption{Evolution of total disc mass for our set of \ACE\ runs with different belt eccentricities $e\sbs{b}$. The spread in orbital eccentricities is set to be equal in all four runs, resulting in nearly identical mass evolution. Deviations are caused by limited grid resolution.
  \label{fig:MassEvolution}}
\end{figure}

For the large parent bodies that are unaffected by radiation pressure, a non-vanishing \emph{mean} eccentricity vector $(e\cos\varpi, e\sin\varpi)$ corresponds to a global offset. The typical rates and velocities at which collisions occur are dictated by the \emph{spread} in orbital elements. Discs with different mean eccentricities will still have similar erosion rates as long as this spread is comparable. The parallel evolution of total mass and dust mass in our simulation runs for different $e\sbs{b}$, shown in Figure~\ref{fig:MassEvolution}, confirms this expectation. When looking at parent bodies and larger grains, a global eccentricity results in literally just an offset, where the relative widths of periastron and apastron side are equal.

\begin{figure}
  \centering
  \includegraphics[width=\hsize]{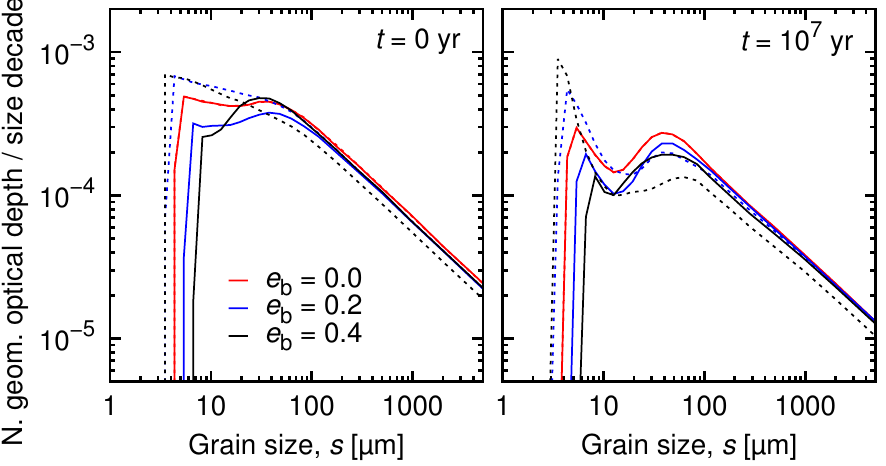}
  \caption{Grain size distributions for a set of three discs with different belt eccentricities $e\sbs{b}$ at \textsl{(left)} $t \approx 0$~yr and \textsl{(right)} $t = 10^7$~yr. Solid lines and dashed lines trace the distributions at the pericentres and apocentres of the parent belts, respectively.}
  \label{fig:size_dist}
\end{figure}

The picture is different for smaller grains. The additional action of radiation pressure induces a lower blowout limit that depends on the birth location, changing the size distribution in the parent belt. Blowout occurs for smaller values of $\beta$ for grains that are launched near periastron of a parent orbit. The excess velocity there helps them overcome the gravitational bond, increasing the maximum size of unbound grains.
Vice versa, when released near the belt apastron, smaller grains can stay bound. Figure~\ref{fig:size_dist} illustrates this shift in blowout size with grain size distributions near the periastra and apastra of parent belts with different $e\sbs{b}$. According to equation~(\ref{eq:blowout}), the ratio between the blowout sizes at periastron and apastron is given by $(1 + e\sbs{b})/(1 - e\sbs{b})$. This ratio reaches a value of $7/3$ for $e\sbs{b} = 0.4$, which is consistent with the \ACE\ results.

\begin{figure}
  \centering
  \includegraphics[width=\hsize]{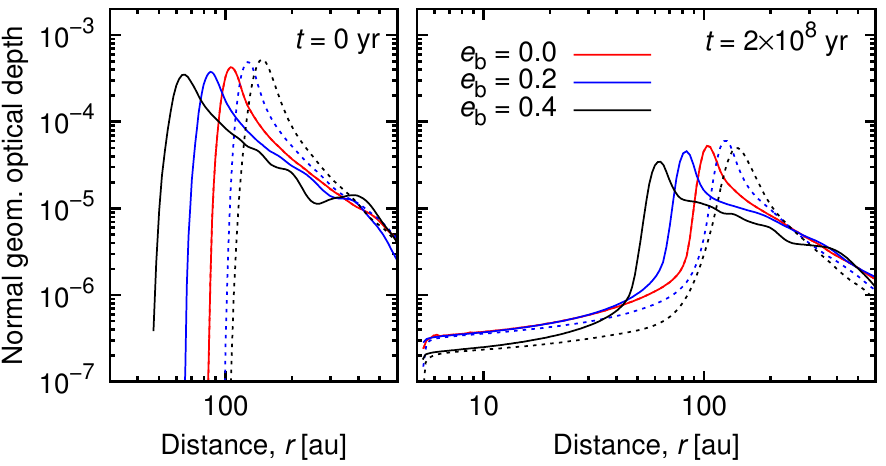}
  \caption{Radial distributions of optical depth for a set of three discs with different global eccentricities $e\sbs{b}$ at \textsl{(left)} $t \approx 0$~yr and \textsl{(right)} $t = 2\times 10^8$~yr. Solid lines and dashed lines follow radial cuts through the periastron and apastron sides of the parent belts, respectively.}
  \label{fig:tau_profs}
\end{figure}

In the left panel of Figure~\ref{fig:tau_profs}, radial profiles of the normal optical depth, $\tau$, are plotted for $t \approx 0$~yr. At large radii, these profiles drop almost as $\tau \propto r^{-3/2}$, the behaviour expected for discs in equilibrium \citep{krivov+2006}. This is by design because we employed the initial setup of \citet{strubbe+chiang2006}, who derived that slope analytically. Wiggles in the radial profiles are artefacts of the narrow size distribution in the halo reaching the resolution limit of our mass grid.

\begin{figure}
  \centering
  \includegraphics[width=\hsize]{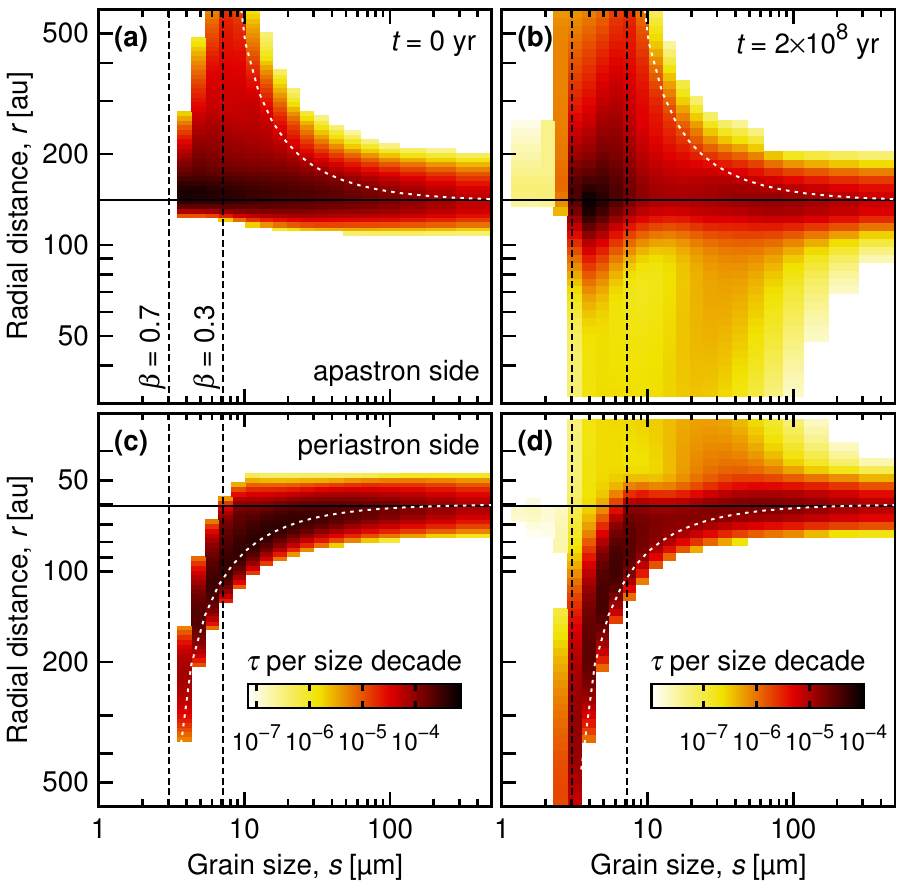}
  \caption{Distribution of grains over sizes and radial distances in a disc with $e\sbs{b} = 0.4$ in two directions:
          \emph{(top)} towards the apastron of the eccentric parent belt and \emph{(bottom)} towards the periastron.
          Colour scale indicates normal optical depth per size decade.
          The dashed vertical lines represents the blowout limit due to radiation pressure
          for release near periastron ($\beta = 0.3$) and apastron ($\beta = 0.7$), respectively.
          The belt of parent bodies is marked with solid horizontal lines.
          Dotted curves trace the apastron distances of grains produced on the respectively opposing sides of the parent belt.
  }
  \label{fig:m_of_s_and_r}
\end{figure}

The blowout limit then defines the typical sizes of the barely bound grains that form the halo. The different blowout limits induce different grain sizes, depending on the side of the halo. Barely bound grains are typically produced near their periastra, but spend most of their time near their orbits' apastra, i.\,e. on the opposite side. Therefore the part of the halo that extends beyond the apastron side of the belt is produced at its periastron, and vice versa. The two upper rows in the left column of Figure~\ref{fig:m_of_e_and_w} show how larger grains have orbits that are still aligned azimuthally with the parent belt. Their periastra (as described by $e$ and $\varpi$ in that figure) remain close to the belt's periastron. The small grains shown in the bottom row of Figure~\ref{fig:m_of_e_and_w} can only stay bound when they are created with their periastra at the apastron side of the belt. In consequence, the halo on the apastron side will be formed by grains larger than those on the periastron side. Figure~\ref{fig:m_of_s_and_r} shows this effect for a disc with $e\sbs{b} = 0.4$. There, the $\beta$ values of the bound halo grains on the periastron side approach $0.7$ as distance increases. The halo on the apastron side is populated by larger grains, initially limited by radiation pressure to $\beta < 0.3$.

\begin{figure}
  \centering
  \includegraphics[width=\hsize]{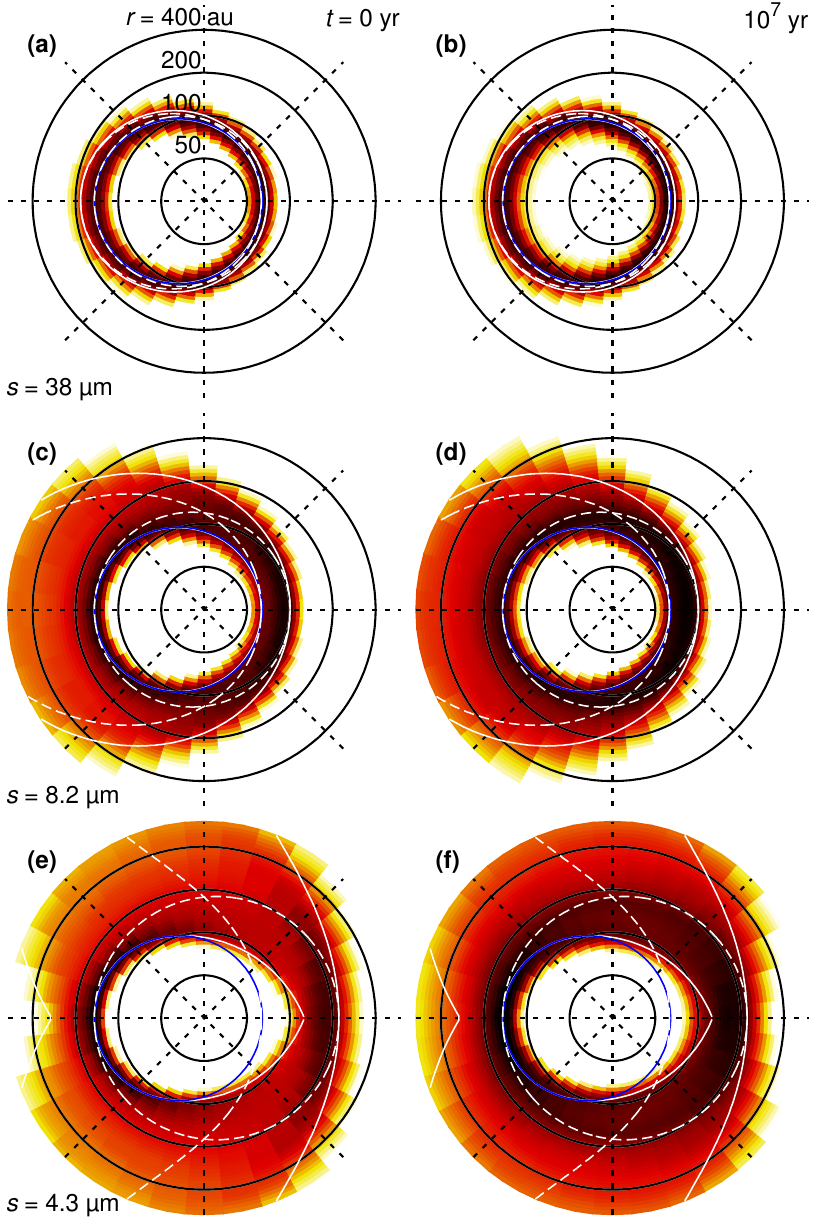}
  \caption{Two-dimensional distributions of grains for the same grain sizes and stages as in Figure~\ref{fig:m_of_e_and_w}.
          Radial distance is scaled logarithmically, with black circles indicating equal distances,
          spaced at factors of two. Colour scale indicates normal optical depth per size decade, spanning 2.5 orders of magnitude from white to black.
          Solid blue lines trace the belts of parent bodies. Dashed white lines follow the trajectories
          of fragments launched from the parent belts' periastra and apastra.
          Solid white caustics separate regions that can be reached by bound grains directly originating in a thin parent belt from regions that cannot.
          Only for the barely bound grains in the bottom row does the inner caustic differ from the parent belt itself.
  }
  \label{fig:m_of_r_and_w}
\end{figure}

The widths of the size distributions on the two sides differ as illustrated in Figure~\ref{fig:m_of_r_and_w}, where radial and azimuthal distributions for different grain sizes are compared. For all three grain sizes shown, the range of distances covered beyond the apastron side of the belt is wider because they are bound less in that direction. The regions populated by grains of different sizes overlap more strongly on the apastron side, and hence, grains of a wider range of sizes populate a given distance.

The main reason for the asymmetries seen in the halos in our simulations are the different conditions under which grains are launched from different sites along the parent belt. Initial orbital velocities differ along the belt, and so do the resulting fragment orbits. Most of the effects described in this section can also be found in \citet{lee+chiang2016}. A prominent example is the bow or wing asymmetry seen in Figure~\ref{fig:m_of_r_and_w}e as a dark arc on the right-hand side of that panel. \citeauthor{lee+chiang2016} have an equivalent arc in their Figure~1 (and Figure~7). It traces the outer boundary of the region populated by barely bound grains on the periastron side.

A similarly asymmetric halo is described by \citet{kral+2015} for dust produced in a giant breakup. However, the asymmetry there is mainly caused by the asymmetric distribution of launch sites, with the initial round of fragments all being produced from a single parent.

\subsection{Effects of collisions}\label{sec:Distributions:Collisions}
In collisional equilibrium a size distribution of infinite extent can be well-described by a power law $n(s) \propto s^\alpha$, with $\alpha \sim -3.5$ \citep{dohnanyi1969,durda+dermott1997,o'brien+greenberg2003,wyatt+2011,pan+schlichting2012}. However, notable ripples appear near physical breaks or cutoffs. Waves in the size distribution are induced for asteroids by the transition from strength to self-gravity \citep{durda+dermott1997} and for grains above the blowout limit by the radiation pressure cutoff \citep{campo-bagatin+1994a}. At this lower size end, barely bound grains become overabundant because of a lack of smaller projectiles. In turn, this overabundance leads to a depletion of somewhat larger grains. Wavelengths and amplitudes of these waves are determined by impact energies relative to disruption thresholds \citep{krivov+2006}. More realistic impact physics, such as cratering collisions, quickly damp the waves towards larger grain sizes \citep{thebault+2003,thebault+augereau2007,mueller+2010}, when compared to simulations where only disruptive collisions are considered \citep{loehne+2008}.

This wave near the blowout limit is overlaid by an effect first described in \citet{thebault+wu2008}. For larger grains, the typical orbital eccentricities, and hence the typical relative velocities and collision timescales, are determined by that of their parent bodies. For smaller grains, radiation pressure is more important. An additional break in the size distribution occurs where the two effects are equal, i.\,e. where $e$ from equation~(\ref{eqResult_e}) equals $e\sbs{b}$. For grains produced in a parent belt with average proper eccentricities $e\sbs{p} = 0.05$, this break is expected near $\beta = e\sbs{p}/(1 + e\sbs{p}) \approx 0.05$, corresponding to grain sizes $s \approx 45$~\micron\ in our setup. The right panel of Figure~\ref{fig:size_dist} shows both the break around this size and the depletion and blowout-induced waviness below.

The closer grain sizes get to the blowout limit, the further do size distributions near the parent-belt apastra and periastra deviate from one another. The peak just above the blowout limit is higher at the parent apastra than at the parent periastra. This difference increases with increasing $e\sbs{b}$, reaching an order of magnitude for $e\sbs{b} = 0.4$ (Fig.~\ref{fig:size_dist}). This can be understood because the two sides are coupled; the larger grains coming from the belts' periastra suffer from collisions with the smaller grains coming from the apastra as their orbits cross.

For grains larger than around 20~\micron, the situation seems reversed. In collisional equilibrium, grains in this size range contribute more to the optical depth at the belt periastra. This effect is mainly caused by these grains being spread out more widely on the apastron side, leading to lower densities there. However, part of this asymmetry extends to grains sizes where radiation pressure and the resulting radial spread are neglible. In consequence, the size distributions are shallower overall for grains between 20~\micron\ and 1~mm, which translates to shallower spectral energy distributions (SEDs) in the corresponding range of wavelengths \citep{draine2006}. When combined with the increased abundance of barely bound grains at the apastra of discs with higher $e\sbs{b}$, the effective grain sizes shift towards smaller radii. This would imply higher temperatures near the apastra, reducing the brightness asymmetry due to pericentre glow at shorter wavelength. The detailed effects on the observable SEDs and images will be discussed in Section~\ref{sec:Observables}.

\begin{figure}
  \centering
  \includegraphics[width=\hsize]{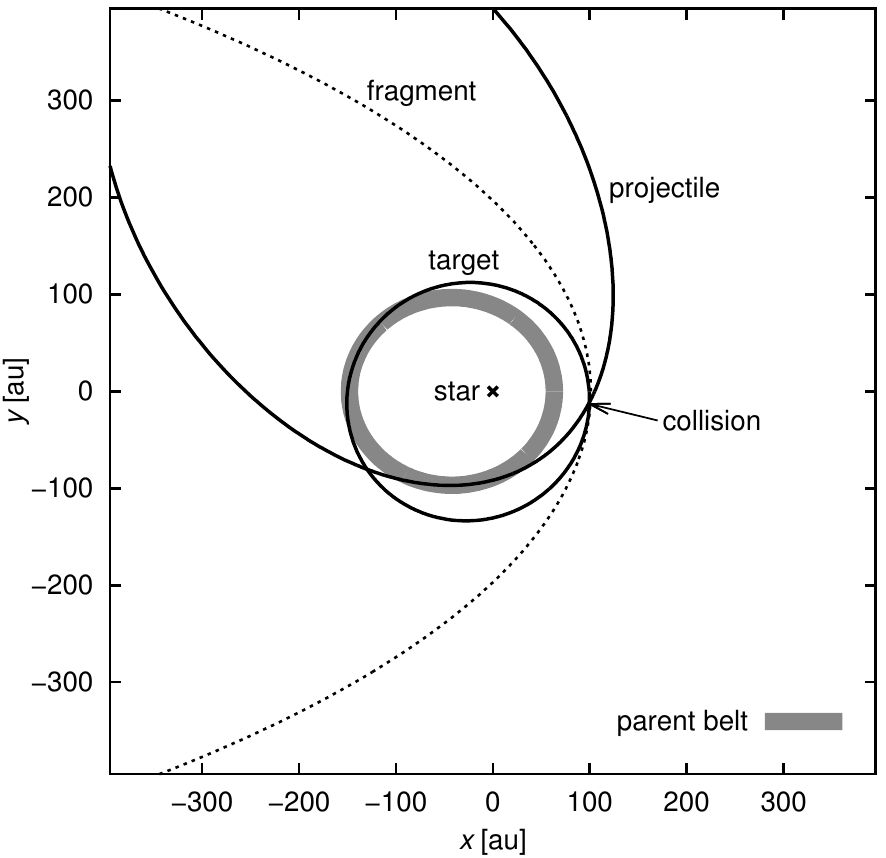}
  \caption{Two grains that are produced in the parent belt collide and launch a barely bound fragment that is apsidally aligned with the belt. See text for details.\label{fig:SampleFragmentOrbit}}
\end{figure}

In Figure~\ref{fig:m_of_e_and_w} the most notable difference between the initial stage and the intermediate stage is the filling of regions that could not be reached initially. In particular, grains with $s = 4.3$~\micron\ ($\beta = 0.5$) appear on bound orbits ($e < 1$) with periastra aligned with the periastron of the belt. These grains cannot be just produced in the parent belt. P--R drag cannot be responsible either because it cannot alter $\varpi$ and it cannot re-bind unbound grains in significant numbers (which is why this process is not modelled in \ACE). Thus, they must stem from medium sized grains. An illustration of this process is given in Figure~\ref{fig:SampleFragmentOrbit}, where the orbits of three particle types actually represented in the \ACE\ grid are shown. The target ($s = 12$~\micron, $\beta = 0.17$, $q = 98$~au, $e = 0.22$, $\varpi = 23^\circ$) and the projectile ($s = 5.4$~\micron, $\beta = 0.4$, $q = 85$~au, $e = 0.75$, $\varpi = -56^\circ$) are produced at different locations in the parent belt. At one of their mutual collision points, they produce a barely bound fragment that is aligned with the parent belt ($s = 4.3$~\micron, $\beta = 0.5$, $q = 101$~au, $e = 0.95$, $\varpi = 0^\circ$). This symmetrization of the phase space of grains with $0.3 < \beta < 0.7$ leads to a symmetrization of their spatial distribution. The arc seen on the periastron side in Figure~\ref{fig:m_of_r_and_w}e is gone in Figure~\ref{fig:m_of_r_and_w}f. Figure~\ref{fig:m_of_s_and_r}b shows how these grains then contribute to and strengthen the halo on the apastron side.

Comparison of the panels in Figure~\ref{fig:tau_profs} suggests that the halo on the periastron side is strengthened even more. While the circular belt produces a halo with a classical $\tau \propto r^{-3/2}$ behaviour \citep{strubbe+chiang2006,krivov+2006}, the halos on the periastron sides of eccentric belts are closer to $\tau \propto r^{-1\ldots-3/4}$. The apastron sides see a steeper falloff. The slopes of all curves converge beyond $\sim 400$~au. 

\subsection{Inward transport through drag forces}\label{sec:Distributions:Drag}
Poynting--Robertson drag and stellar wind drag cause grains to spiral towards the star on timescales set by their size-dependent susceptibility to radiation and wind pressure, stellar luminosity and mass loss rate, and the grains' orbital semi-major axes and eccentricities \citep[see, e.\,g,][]{robertson1937,wyatt+whipple1950,burns+1979}. The classical results for the orbit-averaged reduction rates of $a$s and $e$s are
\begin{eqnarray}
  \dot a &=& -\frac{\beta GM_*}{ca} \frac{2 + 3 e^2}{(1 - e^2)^{3/2}},\label{eq:a_dot}\\
  \dot e &=& -\frac{5\beta GM_*}{2ca^2} \frac{e}{(1 - e^2)^{1/2}}.\label{eq:e_dot}
\end{eqnarray}
Other orbital elements are not affected secularly and non-relativistically. When grains are released from circular orbits in a source belt, drag alone produces constant optical depth $\tau(r)$ towards the star. Collisional sinks \citep{wyatt+1999,wyatt2005a} make $\tau$ decrease further in. Additional sources, such as active comets \citep{leinert+1983}, increase $\tau$.

Figure~\ref{fig:tau_profs} shows how drag and collisions shape the optical depth profiles in the inner regions of our model runs. From the peak in the parent belt to its inner edge, optical depth drops by about 2 orders of magnitude. Closer to the star, the slope flattens out as collisions become less important.

In the runs with eccentric parent belts, optical depths differ between apastron and periastron sides. Values on the apastron sides are systematically lower because drag rates are higher there.
This can be explained in more detail with the following analytic model. At a given distance $r$ from the star, the optical depth is determined by three factors: (1) the rate $\dot{\sigma}$ at which cross section gets dragged accross that distance, (2) the azimuthal spread of that cross section, caused by orbital speed $v$, and (3) the radial spread, caused by radial drift speed $\dot{r}$ times orbital period $P$. All these quantities differ between periastron and apastron side. On the periastron side, we have
\begin{equation}
  \tau_q = \frac{\dot\sigma_q(r)}{|\dot r_q| P_q v_q},\label{eq:tau_q}
\end{equation}
and for $r = q = a_q (1-e_q)$, the product of orbital period and orbital speed is
\begin{equation}
  P_q v_q = 2\pi a_q \sqrt{2a_q/r - 1} = \frac{2\pi r}{1 - e_q} \sqrt{\frac{1 + e_q}{1 - e_q}}.\label{eq:Pv_q}
\end{equation}
The radial component of the orbital velocity vanishes and $\dot r$ is given by the P--R induced reduction of periastron distance:
\begin{equation}
  \dot{r}_q = \dot q = \dot a_q (1 - e_q) - a_q \dot e_q.\label{eq:q_dot_raw}
\end{equation}
Inserting equations~(\ref{eq:a_dot}) and (\ref{eq:e_dot}) into (\ref{eq:q_dot_raw})
results in
\begin{equation}
  \dot{q} = -\frac{\beta GM_*}{2cq} \frac{4 + e_q^2 - 5e_q}{1 + e_q} \sqrt{\frac{1 - e_q}{1 + e_q}}.\label{eq:q_dot}
\end{equation}
and
\begin{equation}
  |\dot{r}_q| P_q v_q = \frac{\beta \pi GM_*}{c} \frac{4 + e_q^2 - 5e_q}{1 - e_q^2}.\label{eq:rPv_q}
\end{equation}
At apastron, where $r = Q = a_Q (1 + e_Q)$, we find
\begin{equation}
  |\dot{r}_Q| P_Q v_Q = \frac{\beta \pi GM_*}{c} \frac{4 + e_Q^2 + 5e_Q}{1 - e_Q^2}.\label{eq:rPv_Q}
\end{equation}
For the flux of cross section, we assume
\begin{equation}
  \dot \sigma_Q\left[r = Q\right] = \dot \sigma_a\left[\frac{Q}{1 + e_Q}\right] = \dot \sigma_q\left[Q\frac{1 - e_Q}{1 + e_Q}\right],
\end{equation}
i.\,e. material flux does not change from periastron to apastron of a single orbit.
The ratio of fluxes at the same distance on both sides is thus given by
\begin{equation}
  \frac{\dot \sigma_Q(r)}{\dot \sigma_q(r)} = \frac{\dot \sigma_a\left[r/(1 + e_Q)\right]}{\dot \sigma_a\left[r/(1 - e_q)\right]}.\label{eq:sigma_ratio}
\end{equation}
For the radial dependence of $\dot\sigma_a(r)$ under the action of drag and collisions, we adopt the analytical result of \citet{wyatt+1999}:
\begin{equation}
  \dot\sigma_a(r) \propto \left[1 + 4\eta(1 - \sqrt{r/r_0})\right]^{-1},\label{eq:sigma_wyatt}
\end{equation}
where $r_0 = a\sbs{b} = 100$~au and $\eta \approx 100$ is the ratio of drag and collision time scales in the parent belt.

The combination of equations~(\ref{eq:tau_q}), (\ref{eq:rPv_q}), (\ref{eq:rPv_Q}), (\ref{eq:sigma_ratio}), and (\ref{eq:sigma_wyatt}) leads to
\begin{equation}
  \frac{\tau_Q}{\tau_q} = \frac{1 + 4\eta\left(1 - \sqrt{r/[r_0(1 + e_Q)]}\right)}{1 + 4\eta\left(1 - \sqrt{r/[r_0(1 - e_q)]}\right)} \frac{4 + e_q^2 - 5e_q}{4 + e_Q^2 + 5e_Q} \frac{1 - e_Q^2}{1 - e_q^2}.
\end{equation}
The first term on the right-hand side accounts for collisional loss. It dominates close to the parent belt and for high $\eta$. The remainder describes pure drag, dominating close to the star and for low $\eta$.

P--R drag reduces orbital eccentricities as grains spiral in. At a given distance $r$ from the star, grains on the apastron side will therefore have an eccentricity $e_Q(r)$ that is lower than the corresponding eccentricity $e_q(r)$ of grains on the periastron side because $r$ is reached later on the apastron side. The exact relation between the two eccentricities can be deduced from integrating the evolutions of $a$ and $e$ simultaneously. From
\begin{equation}
  \frac{\total e}{\total a} = \frac{\dot e}{\dot a} = \frac{5e(1-e^2)}{2a(2 + 3e^2)},
\end{equation}
\citet{wyatt+whipple1950} obtain
\begin{equation}
  \frac{a_2}{a_1} = \frac{1 - e_1^2}{1 - e_2^2} \left(\frac{e_2}{e_1}\right)^{4/5}
\end{equation}
for P--R drag between States 1 and 2.
For $q = q_1$, $Q = Q_2$, and $r = q = Q$ this leads to
\begin{equation}
  \frac{Q_2}{q_1} = \frac{a_2}{a_1} \frac{1 + e_2}{1 - e_1} = \frac{1 + e_1}{1 - e_2} \left(\frac{e_2}{e_1}\right)^{4/5} = 1,
\end{equation}
which can be solved numerically to find $e_Q(= e_2)$ as a function of $e_q(= e_1)$.

\begin{figure}
  \centering
  \includegraphics[width=\hsize]{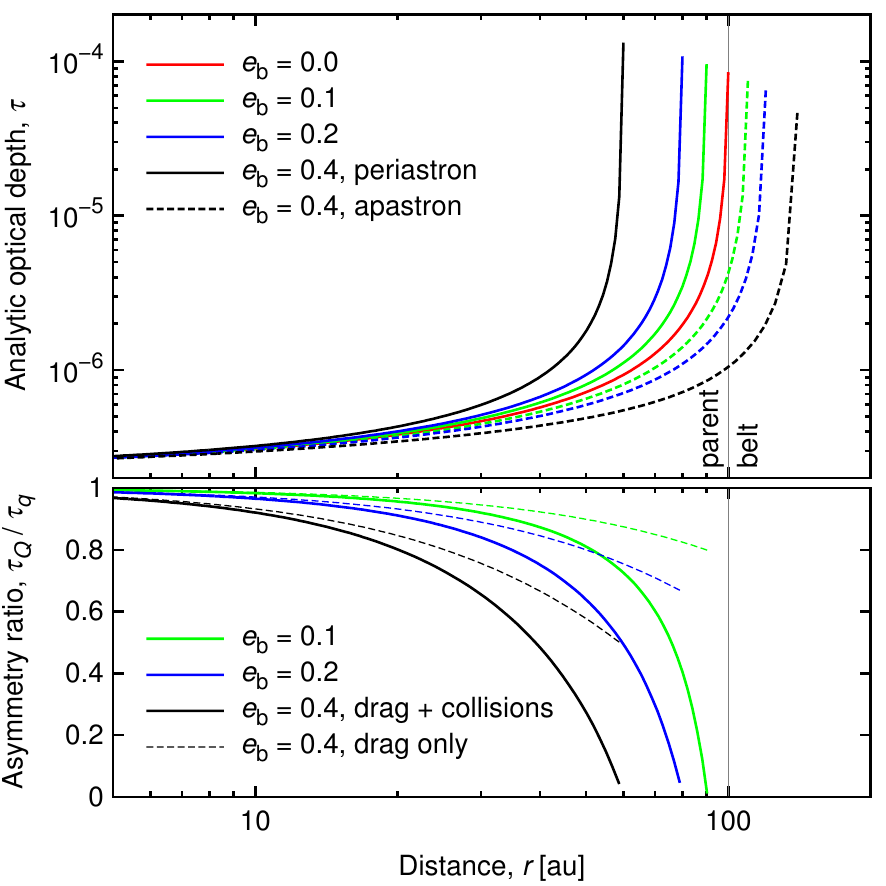}
  \caption{\emph{Top:} Analytically derived radial profiles of normal optical depths on the \emph{(dashed lines)} apastron and the \emph{(solid lines)} periastron sides as a function of distance. Grains are assumed to originate from belts with $r_0 = a\sbs{b} = 100$~au and eccentricities $e\sbs{b}$ of \emph{(red)}  $0.0$, \emph{(blue)} $0.2$, and \emph{(black)} $0.4$. The ratio of drag and collision time scales is set to $\eta = 100$. At the belt edges, i.\,e. at the right ends of the curves, optical depth starts at finite values. \emph{Bottom:} Ratios of optical depths on opposing sides for the same set of parameters. Dashed lines show the contribution from drag alone.}
  \label{fig:TauProfsAnalytic}
\end{figure}

In Figure~\ref{fig:TauProfsAnalytic}, we show the profiles of optical depth expected from this analytic approach. Comparison with Figure~\ref{fig:tau_profs} shows that these profiles can well reproduce the results of our \ACE\ runs outside of the parent region, i.\,e. for $r < q\sbs{min} = (a\sbs{b} - \Delta a\sbs{b}) (1 - e\sbs{b} - \Delta e\sbs{b})$ on the periastron side and $r < Q\sbs{min} = (a\sbs{b} - \Delta a\sbs{b}) (1 + e\sbs{b} - \Delta e\sbs{b})$ on the apastron side. The lower panel of Figure~\ref{fig:TauProfsAnalytic} shows the resulting asymmetry ratios for a range of eccentricities $e_b$ and distances $r$. 
For $e_b = 0.2$ and $r = 0.5 a\sbs{b}$, we find $\tau\sbs{Q}/\tau\sbs{q} = 0.6$. The asymmetry is vanishing slowly with decreasing distance from the star because typical eccentricities also decrease. This trend is seen both in the \ACE\ output and the analytic model.

For $e = e\sbs{b} \approx 0.1$ observed for the outer Fomalhaut disc, we predict a flux deficit of 20\,\% at $r = 0.5 a\sbs{b} \approx 70$~au on the apastron side compared to the periastron side. This asymmetry is not seen in currently available observational data because these are limited by either low resolution in the case of \textit{Herschel}/PACS \citep{acke+2012} or sensitivity in the cases of HST \citep{kalas+2005} and ALMA \citep{boley+2012}. If the dust in the region interior to the outer Fomalhaut belt does not exhibit such an asymmetry, this would speak against inward drag as the dominating mechanism for replenishment.

Note that the above analysis assumes that grains in the P--R region start from orbits that follow those of the parent bodies, i.\,e. they are large enough not to be affected strongly by radiation pressure. However, Figure~\ref{fig:m_of_s_and_r} shows that smaller grains with potentially higher initial eccentricities contribute as well. A higher typical eccentricity of P--R grains would further strengthen the asymmetry. Despite uncertainties in $e$ and the simplified collisional depletion, the effect is robust and provides a testable prediction for the asymmetry of the drag-filled inner regions of eccentric belts.

\subsection{Comparison with precession timescales}\label{sec:Distributions:Timescales}
In the presented \ACE\ runs we assumed a fixed average belt eccentricity and orientation, both present from the beginning and static throughout the simulation. At the same time we show in Figure~\ref{fig:Lifetimes} that a modest Neptune at 30~au can already induce precession periods as short as $\sim 10^8$~yr for a distant belt at 100~au. In the regime of observable dust, precession can thus act on timescales longer than those for collisions, but shorter than those of drag. If precession periods and forced eccentricities were equal for all objects, the effects on dust distribution would be negligible. The disc would precess as a whole while collisions and drag take place. However, differential precession due to different semi-major axes and $\beta$ ratios will twist the disc. Dragged-in dust will precess faster, dust in the halo slower. In regions where collision timescales are longer, precession will smear the distribution of complex eccentricities, increasing the spread in $\Delta e$, collision velocities, and potentially, depletion rates. An updated model that accounts for this process is in preparation.

\section{Spectral energy distributions and images}\label{sec:Observables}
\begin{figure}
  \centering
  \includegraphics[width=\hsize]{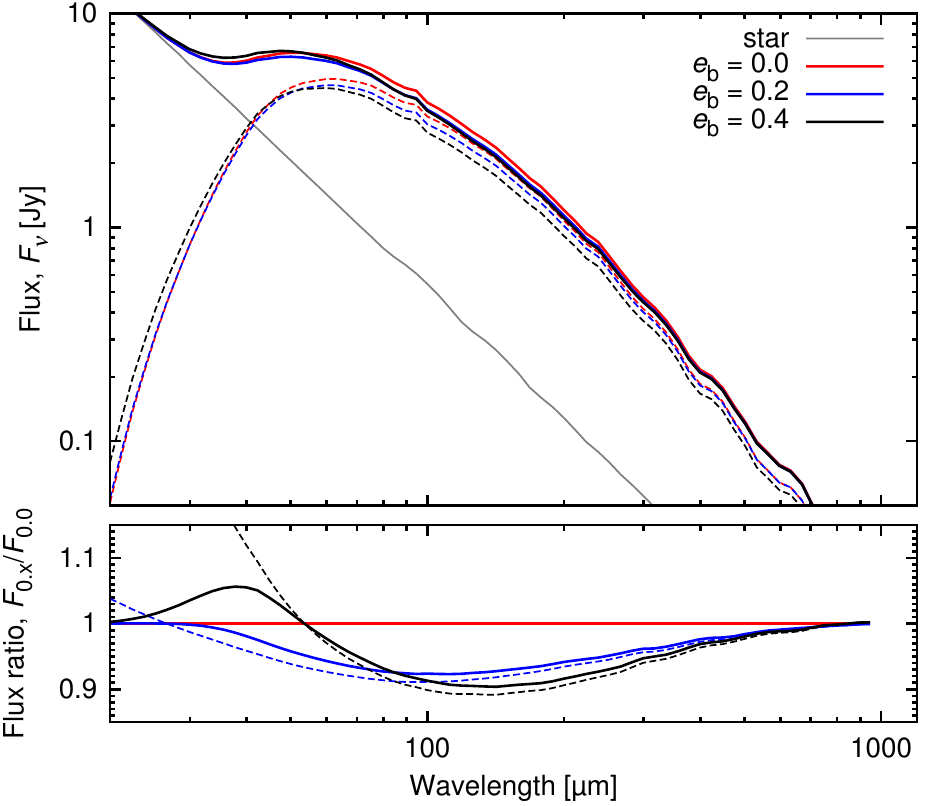}
  \caption{Spectral energy distributions at $t = 2 \times 10^8$ yr for \textsl{(red)} $e\sbs{b} = 0.0$, \textsl{(blue)} $e\sbs{b} = 0.2$, and \textsl{(black)} $e\sbs{b} = 0.4$. Solid lines stand for the combined emission from star and disc, dashed lines represent the disc alone. In the bottom panel, flux ratios relative to the disc with $e\sbs{b} = 0.0$ are plotted. Fluxes are scaled marginally such that they converge at a wavelength of 1\,mm.}
  \label{fig:SEDs}
\end{figure}

Figure~\ref{fig:SEDs} illustrates the mild influence that the combination of geometrical offset, collisions and drag has on the overall SED. Even for the rather eccentric cases $e\sbs{b} = 0.2$ and $s\sbs{b} = 0.4$, the differences in the wavelength range above a few tens of microns do not exceed 10\,\%. It is only at shorter wavelengths that fluxes from eccentric discs become significantly higher, which is mainly due to pericentre glow, i.\,e. grains near the belt periastron having higher temperatures.

At wavelengths $\lambda \gtrsim 100$~\micron, the SEDs reflect the differences in size distributions described in Section~\ref{sec:Distributions:Collisions}. Shallower size distributions in more eccentric belts result in shallower SEDs. The discrepancy of 10\,\% over a factor of ten in wavelength for $e\sbs{b} = 0.4$ corresponds to a difference in power-law slopes of $\log(1.1)/\log(10) \approx 0.04$. There exist several effects that have stronger impacts on SED slopes, but an unknown eccentricity adds to the uncertainty in the derivation of these other parameters. SED slopes are commonly used to infer underlying grain size distributions, which in turn are related to, e.\,g., collisional physics. Assuming that a difference of $0.04$ in the SED slopes translates to no more than $0.04$ in the inferred slopes of the size distributions \citep[cf.][]{draine2006}, we conclude that other observational and modelling uncertainties dominate.

\begin{figure*}
  \centering
  \includegraphics{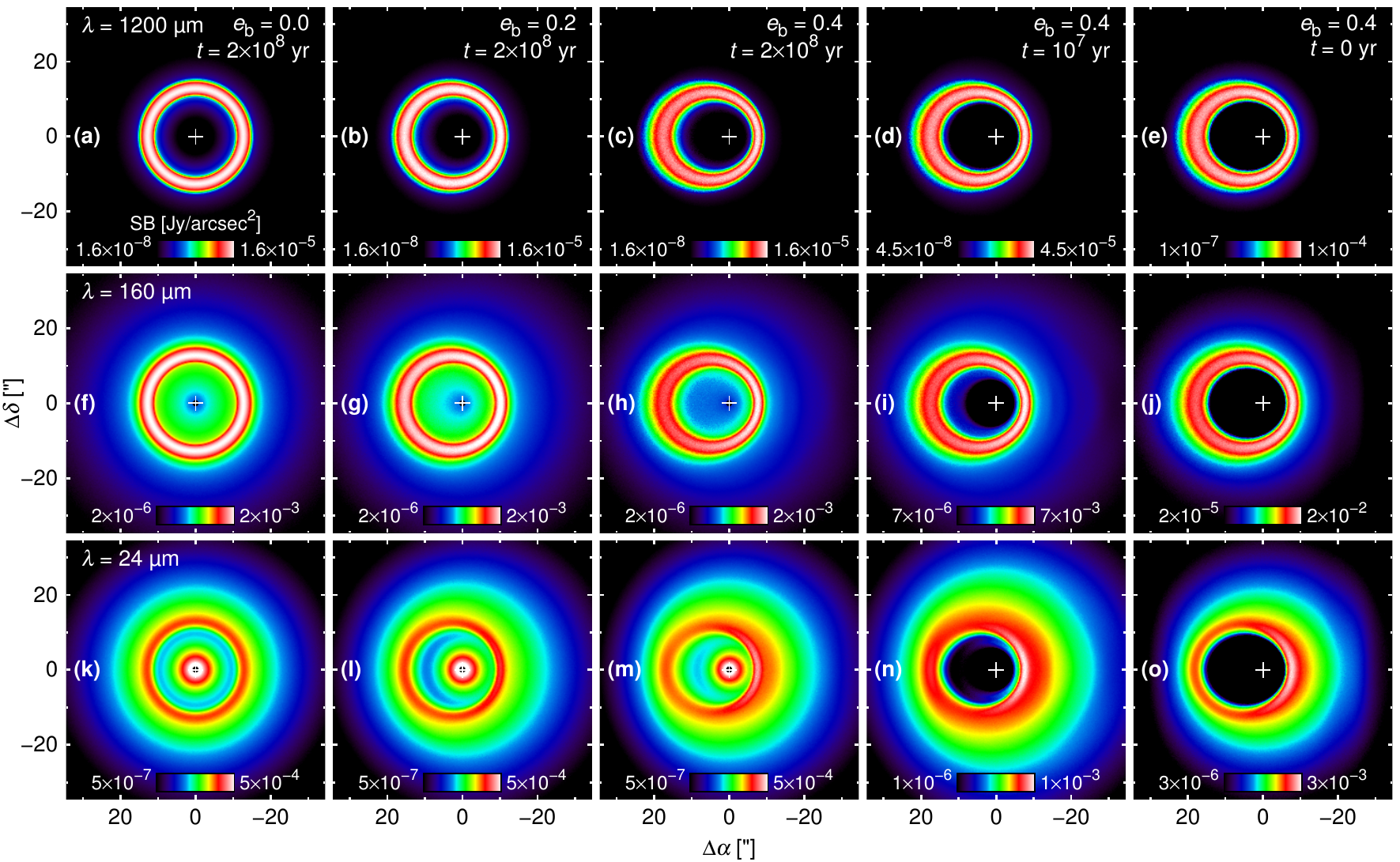}
  \caption{Synthetic maps of surface brightness at \textsl{(top)} $\lambda = 1200$~\micron, \textsl{(middle)} 160~\micron, and \textsl{(bottom)} 24~\micron. Underlying discs are seen face-on, from a distance of 8~pc. The first three columns from left show discs with belt eccentricities $e\sbs{b} = 0.0$, $0.2$, and $0.4$ at time $t = 2\times 10^8$~yr. Columns 4 and 5 show the disc with $e\sbs{b} = 0.4$ at the intermediate ($t = 10^7$~yr) and initial ($t = 0$~yr) stage. Colour scales indicate logarithm of surface brightness.
  \label{fig:Images}}
\end{figure*}

The panels in Figure~\ref{fig:Images} show the fiducial discs with belt eccentricities $e\sbs{b} = 0.0$, $0.2$, and $0.4$ in thermal emission at 24~\micron, 160~\micron, and 1.2~mm. Characteristic grain sizes $s\sbs{c}$ at different wavelengths $\lambda$ can be estimated from $s\sbs{c} \approx \lambda/2\pi$ \citep{backman+paresce1993}. At $\lambda = 1.2$~mm, the halo is invisible and the discs appear as narrow belts because the dominant grains have $s\sbs{c} \approx 200$~\micron, and with $\beta \approx 0.01$, are only weakly affected by radiation pressure. The drag timescales are such that these large grains just start to fill the inner gap after a few times $10^8$~yr (Fig.~\ref{fig:Lifetimes}). At 160~\micron, correponding to $s\sbs{c} \approx 25$~\micron\ and $\beta \approx 0.08$, the belt is wider and the halo and the inner region start to become visible. As long as drag is not important, though, the low $\beta$ of the grains observed makes the chosen initial distribution, which corresponds to the setup by \citet{lee+chiang2016}, a good proxy to the collisional steady state at these longer wavelengths.

At $\lambda = 24$~\micron, we find $s\sbs{c} \approx 4$~\micron\ ($\beta \approx 0.5$). Observations at this wavelength are thus sensitive to the distribution of barely bound grains around the A3\,V star assumed in our simulations. Grains of this size only stay bound when released from the apastron side of the parent belt, strengthening the halo on the periastron side. As a result and in contrast to the longer wavelengths, emission on the periastron extends further away from the belt. Images and radial profiles become more symmetric with increasing distance.

\begin{figure}
  \centering
  \includegraphics[width=\hsize]{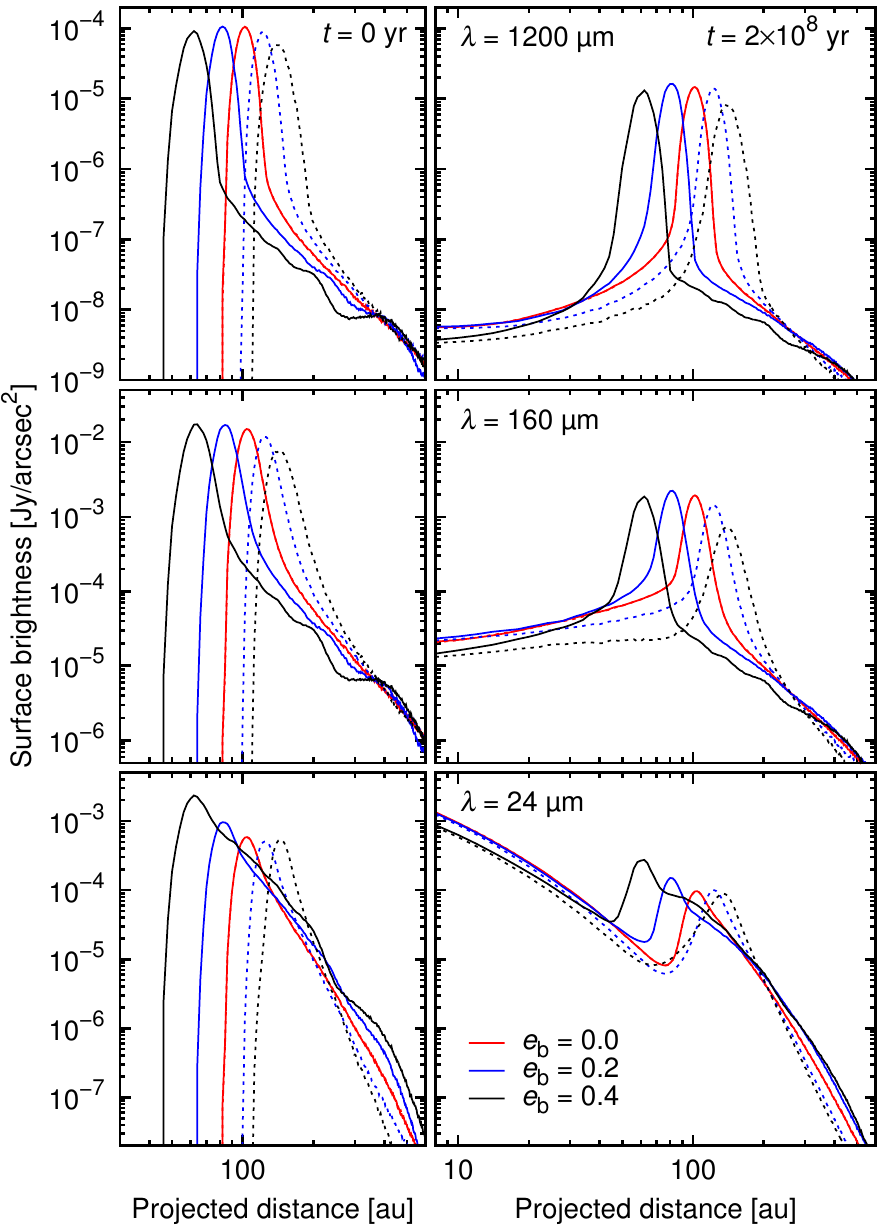}
  \caption{Radial cuts through surface brightness at \textsl{(top)} 1200~\micron, \textsl{(middle)} 160~\micron, and \textsl{(bottom)} 24~\micron. Underlying discs are seen face-on. In the individual panels, the discs with $e\sbs{b} = 0.0$ are plotted in red, $e\sbs{b} = 0.2$ in blue, and $e\sbs{b} = 0.4$ in black. Solid lines trace brightness along the periastron side, dashed lines along the apastron side.}
  \label{fig:ColorProfs}
\end{figure}

At the inner edges of all belts, surface brightness drops by about 1.5--2 orders of magnitude, in agreement with the drops in optical depth and at 160~\micron. However, Figure~\ref{fig:ColorProfs} shows that brightness follows $r^{-2.5\ldots -3}$ at 24~\micron in the drag-filled region, increasing strongly towards the star because of the increasing temperature. Although the profile flattens off further towards the star, where 24~\micron\ is no longer on the Wien side, the behaviour can produce a significant total excess -- if the dust is not intercepted by inner planets \citep[e.\,g.,][]{liou+zook1999,reidemeister+2011}. The wiggles seen at distances $r \gtrsim 200$~au in Figure~\ref{fig:ColorProfs} correspond to the wiggles in the radial profiles of optical depth in Figure~\ref{fig:tau_profs}, which are artefacts of the mass binning.

Pericentre glow \citep{wyatt+1999} and broadening of the belt towards its apastron induce an asymmetry between the peak brightnesses in these two loci. The radial cuts in Figure~\ref{fig:ColorProfs} show ratios between the peaks for $e\sbs{b} = 0.4$ that amount to factors of 1.6 at 1.2~mm, 2.6 at 160~\micron, and 3.1 at 24~\micron. Despite the Wien side of the SED being very sensitive to temperature, the asymmetry is only slightly more pronounced at 24~\micron\ than at 160~\micron\ for two reason: (1) because typical grains near belt apastron are smaller than those near periastron, reducing the temperature difference between both sides; and (2) because the periastron side of the belt is wider at 24~\micron. In the initial disc, where only the second reason applies, the brightness ratio is 4.4 at 24~\micron.

\begin{figure}
  \centering
  \includegraphics[width=\hsize]{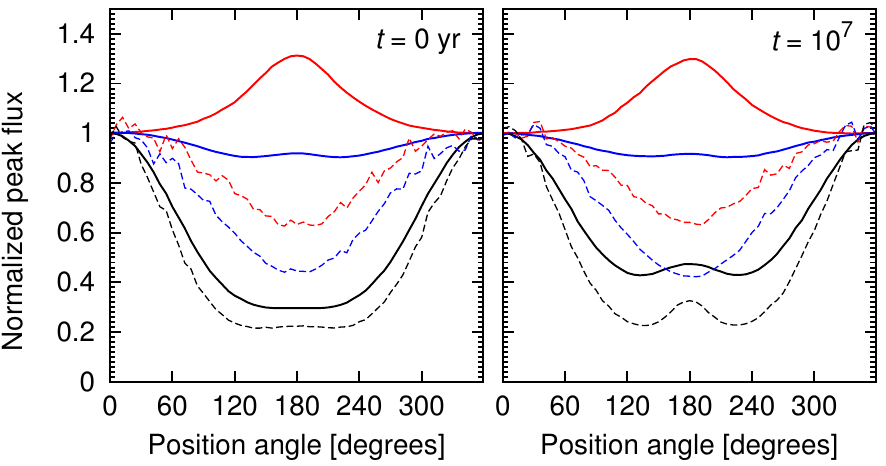}
  \caption{Azimuthal brightness profiles at (left) $t = 0$~yr and (right) $t = 10^7$ yr, taken along radial flux peaks in the images of the belt with $e\sbs{b} = 0.4$. Wavelengths are (top to bottom): 1200~\micron\ (in red), 160~\micron\ (in blue), and 24~\micron\ (in black). Solid lines stand for images PSF-broadened with Gaussians with FWHMs of $8.5''$, $9.7''$, $5.7''$, in the same order. This illustrates apocentre glow at long wavelengths as suggested by \citet{pan+2016}. Dashed lines stand for unsmoothened images.}
  \label{fig:AzimuthalProfiles}
\end{figure}

The actually observed contrast will strongly depend on how well these peaks are resolved. If the narrow periastron side is PSF broadened to the width of the apastron side, the observable difference is drastically reduced. \citet{pan+2016} analyse this effect and show that pericentre glow can turn into apocentre glow at long wavelengths, where thermal emission depends less on temperature and resolution is typically lower. Adopting the idea behind their Figure~2, we smoothened the images with Gaussian kernels and plotted azimuthal profiles for $e\sbs{b} = 0.4$ in our Figure~\ref{fig:AzimuthalProfiles} for two stages. Our initial setup is represented by $t = 0$~yr. After $10^7$~yr, collisional equilibrium is reached but P--R drag has not brightened the innermost region yet. The PSF broadened discs clearly show reduced pericentre glow at 24 and 160~\micron. At 1.2~mm, the apocentre is brighter than the pericentre by a factor of 1.3, roughly consistent with $1 + e\sbs{b}$ derived for small $e\sbs{b}$ by \citet{pan+2016}. The bump seen around 180$^\circ$ at 24~\micron\ on the right panel reflects the tighter radial confinement of the small grains near the belt periastron seen in Figure~\ref{fig:Images}n.

Images in scattered light are similar to 24-\micron\ emission because both trace the small, barely bound grains. The major difference lies in the radial slopes, which are steeper in thermal emission because there the exponential dependence on grain temperature factors in. 

The scattering cross-section and its angular dependence are very sensitive to model assumptions on grain morphology. While Mie theory for perfect spheres works well for $s \ll \lambda$, it fails to reproduce the scattering phase function for irregularly shaped grains with $s \gtrsim \lambda$, where the morphology of surface and sub-surface layers becomes important. Compared to a polished sphere, a rougher surface can increase back-scattering and reduce absorption \citep{pollack+cuzzi1980}. At scattering angles far away from the strong forward diffraction peaks of large grains, the resulting phase functions are flatter. In models of debris discs, these more symmetric phase functions are often associated with the presence of smaller, spherical Mie grains. This degeneracy between small grains and grains with small structures is discussed by \citet{min+2010} and \citet{hedman+stark2015} in their models for the Fomalhaut disc and Saturn's G and D68 rings, respectively.

\begin{figure}
  \centering
  \includegraphics[width=\hsize]{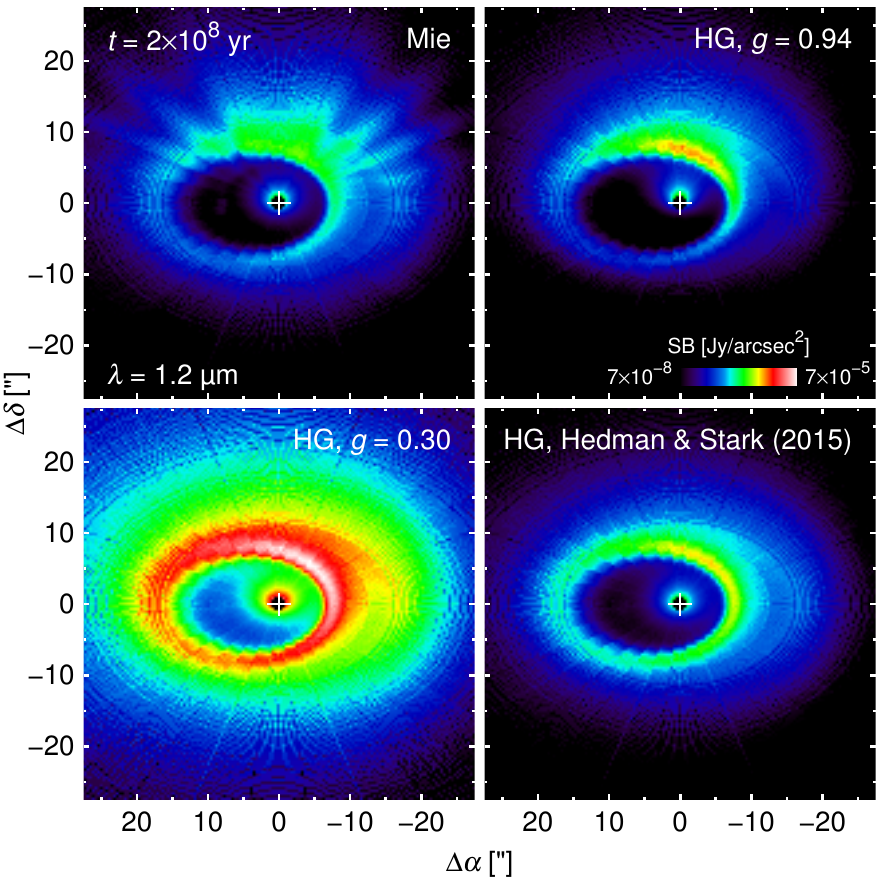}
  \caption{Synthetic scattered light images for different scattering phase functions: \textsl{(top left)} Mie theory for homogeneous spheres, \textsl{(top right)} a Henyey--Greenstein (HG) phase function with anisotropy parameter $g = 0.94$, corresponding to strong forward scattering, \textsl{(bottom left)} HG with mild forward scattering, and \textsl{(bottom right)} three-component HG model of \citet{hedman+stark2015} for Saturn's G ring. The disc is viewed from an angle 45 degrees below its midplane, with the line of nodes aligned east--west and the pericentre due west. Hence, the northern half of the disc lies closer to the observer.}
  \label{fig:PhaseFunction}
\end{figure}

In Figure~\ref{fig:PhaseFunction}, we illustrate this problem with a set of images for different scattering models. Results from Mie calculations are compared to empirical Henyey--Greenstein (HG) phase functions, ranging from mild forward-scattering with anisotropy parameter $g = 0.3$, to strong forward scattering with $g = 0.94$, and the three-component best-fit model that \citet{hedman+stark2015} derive for Saturn's G ring. In the latter, the strongest component has $g = 0.995$. As expected from the rather large grains in our collisional models, the Mie results are best matched by $g$ close to unity. However, the model based on the \citeauthor{hedman+stark2015} fit shows the degree to which the total brightness of the disc in scattered light may be underestimated. Accordingly, non-Mie fits to observed discs find weaker anisotropy, with $g < 0.5$ \citep[e.\,g.,][]{kalas+2005,schneider+2006,debes+2008,thalmann+2011,schneider+2014}.

\section{Conclusions}\label{sec:Summary}
With a new version of our collisional code \ACE, we have studied debris discs that are sustained by eccentric belts of parent bodies. We have identified a set of features and asymmetries that are caused by the combined effects of global eccentricity, radiation pressure, collisional evolution, and drag forces. The features most easily observed are:
\begin{enumerate}
  \item On dynamical timescales, the different radiation pressure blowout limits on opposing disc sides create an asymmetric halo. At shorter wavelengths, where small grains dominate, the halo appears more extended beyond the periastron side of the parent belt. For the larger grains seen at longer wavelengths, the apastron side is more extended.
  \item In collisional equilibrium, the abundance ratios between barely bound grains and grains that are around an order of magnitude larger are different for opposing sides of the belt. On the periastron side, average grains are larger, while on the apastron side, grains are smaller. This size difference reduces the temperature difference between the two sides and weakens the brightness asymmetry expected from pericentre glow.
  \item Poynting--Robertson and stellar wind drag induce an additional asymmetry because they reduce apocentre distances at a higher rate than pericentre distances. Apocentre sides of the drag-filled regions inside of eccentric belts are therefore populated more tenuously. The relative difference between the two sides is comparable with the belt eccentricity, i.\,e. 10\,\% for $e\sbs{b} \approx 0.1$. Towards the star, the asymmetries reduce along with the eccentricities.
  \item Belt eccentricity affects spectral energy distributions only weakly. The azimuthal variation in size distribution and the pericentre glow result in an SED that is broader overall. The effect is most notable in the mid-IR, but significant only for high ($\gtrsim 0.4$) eccentricities.
  \item Interpretation of near-infrared images crucially depends on the scattering model on which it is based. Empirical models valid for the larger grains in the parent belt will be inadequate for the smaller grains that form the outer halo. Mie theory has the benefit of retaining the dependence on grain size, but on the other hand, it does not approximate the scattering phase functions of larger grains well.
\end{enumerate}
A more detailed analysis of the influence of dust optical properties and the disc viewing geometry on the observables is on its way. Another update to \ACE\ will allow us treat the secular precession of orbits by a perturbing planet in parallel with the collisional evolution. 

\begin{acknowledgements}
  We thank an anonymous referee for constructive suggestions that helped improve the presentation.
  Part of this work was supported by the Deutsche Forschungsgemeinschaft through grants LO 1715/2-1, KR 2164/13-1, KR 2164/15-1, WO 857/13-1, and WO 857/15-1.
\end{acknowledgements}


\begin{thebibliography}{78}
\expandafter\ifx\csname natexlab\endcsname\relax\def\natexlab#1{#1}\fi

\bibitem[{{Acke} {et~al.}(2012){Acke}, {Min}, {Dominik}, {Vandenbussche},
  {Sibthorpe}, {Waelkens}, {Olofsson}, {Degroote}, {Smolders}, {Pantin},
  {Barlow}, {Blommaert}, {Brandeker}, {De Meester}, {Dent}, {Exter}, {Di
  Francesco}, {Fridlund}, {Gear}, {Glauser}, {Greaves}, {Harvey}, {Henning},
  {Hogerheijde}, {Holland}, {Huygen}, {Ivison}, {Jean}, {Liseau}, {Naylor},
  {Pilbratt}, {Polehampton}, {Regibo}, {Royer}, {Sicilia-Aguilar}, \&
  {Swinyard}}]{acke+2012}
{Acke}, B., {Min}, M., {Dominik}, C., {et~al.} 2012, \aap, 540, A125

\bibitem[{{Artymowicz} \& {Clampin}(1997)}]{artymowicz+clampin1997}
{Artymowicz}, P. \& {Clampin}, M. 1997, \apj, 490, 863

\bibitem[{{Backman} \& {Paresce}(1993)}]{backman+paresce1993}
{Backman}, D.~E. \& {Paresce}, F. 1993, in Protostars and Planets III, ed.
  E.~H. {Levy} \& J.~I. {Lunine}, 1253--1304

\bibitem[{{Benz} \& {Asphaug}(1999)}]{benz+asphaug1999}
{Benz}, W. \& {Asphaug}, E. 1999, Icarus, 142, 5

\bibitem[{{Beust} {et~al.}(2014){Beust}, {Augereau}, {Bonsor}, {Graham},
  {Kalas}, {Lebreton}, {Lagrange}, {Ertel}, {Faramaz}, \&
  {Th{\'e}bault}}]{beust+2014}
{Beust}, H., {Augereau}, J.-C., {Bonsor}, A., {et~al.} 2014, \aap, 561, A43

\bibitem[{{Boley} {et~al.}(2012){Boley}, {Payne}, \& {Ford}}]{boley+2012}
{Boley}, A.~C., {Payne}, M.~J., \& {Ford}, E.~B. 2012, \apj, 754, 57

\bibitem[{{Bruggeman}(1935)}]{bruggeman1935}
{Bruggeman}, D.~A.~G. 1935, Annalen der Physik, 416, 636

\bibitem[{{Burns} {et~al.}(1979){Burns}, {Lamy}, \& {Soter}}]{burns+1979}
{Burns}, J.~A., {Lamy}, P.~L., \& {Soter}, S. 1979, Icarus, 40, 1

\bibitem[{{Campo Bagatin} {et~al.}(1994){Campo Bagatin}, {Cellino}, {Davis},
  {Farinella}, \& {Paolicchi}}]{campo-bagatin+1994a}
{Campo Bagatin}, A., {Cellino}, A., {Davis}, D.~R., {Farinella}, P., \&
  {Paolicchi}, P. 1994, \planss, 42, 1079

\bibitem[{{Debes} {et~al.}(2009){Debes}, {Weinberger}, \&
  {Kuchner}}]{debes+2009}
{Debes}, J.~H., {Weinberger}, A.~J., \& {Kuchner}, M.~J. 2009, \apj, 702, 318

\bibitem[{{Debes} {et~al.}(2008){Debes}, {Weinberger}, \&
  {Schneider}}]{debes+2008}
{Debes}, J.~H., {Weinberger}, A.~J., \& {Schneider}, G. 2008, \apjl, 673, L191

\bibitem[{{Dohnanyi}(1969)}]{dohnanyi1969}
{Dohnanyi}, J.~S. 1969, \jgr, 74, 2531

\bibitem[{{Draine}(2003)}]{draine2003a}
{Draine}, B.~T. 2003, \apj, 598, 1017

\bibitem[{{Draine}(2006)}]{draine2006}
{Draine}, B.~T. 2006, \apj, 636, 1114

\bibitem[{{Durda} \& {Dermott}(1997)}]{durda+dermott1997}
{Durda}, D.~D. \& {Dermott}, S.~F. 1997, Icarus, 130, 140

\bibitem[{{Esposito} {et~al.}(2016){Esposito}, {Fitzgerald}, {Graham}, {Kalas},
  {Lee}, {Chiang}, {Duch{\^e}ne}, {Wang}, {Millar-Blanchaer}, {Nielsen},
  {Ammons}, {Bruzzone}, {De Rosa}, {Draper}, {Macintosh}, {Marchis}, {Metchev},
  {Perrin}, {Pueyo}, {Rajan}, {Rantakyr{\"o}}, {Vega}, \&
  {Wolff}}]{esposito+2016}
{Esposito}, T.~M., {Fitzgerald}, M.~P., {Graham}, J.~R., {et~al.} 2016, \aj,
  152, 85

\bibitem[{{Grigorieva} {et~al.}(2007){Grigorieva}, {Artymowicz}, \&
  {Th{\'e}bault}}]{grigorieva+2007a}
{Grigorieva}, A., {Artymowicz}, P., \& {Th{\'e}bault}, P. 2007, \aap, 461, 537

\bibitem[{{Hauschildt} {et~al.}(1999){Hauschildt}, {Allard}, \&
  {Baron}}]{hauschildt+1999}
{Hauschildt}, P.~H., {Allard}, F., \& {Baron}, E. 1999, \apj, 512, 377

\bibitem[{{Hedman} \& {Stark}(2015)}]{hedman+stark2015}
{Hedman}, M.~M. \& {Stark}, C.~C. 2015, \apj, 811, 67

\bibitem[{{Hines} {et~al.}(2007){Hines}, {Schneider}, {Hollenbach}, {Mamajek},
  {Hillenbrand}, {Metchev}, {Meyer}, {Carpenter}, {Moro-Mart{\'{\i}}n},
  {Silverstone}, {Kim}, {Henning}, {Bouwman}, \& {Wolf}}]{hines+2007}
{Hines}, D.~C., {Schneider}, G., {Hollenbach}, D., {et~al.} 2007, \apjl, 671,
  L165

\bibitem[{{Hirayama}(1918)}]{hirayama1918}
{Hirayama}, K. 1918, \aj, 31, 185

\bibitem[{{Kalas}(2005)}]{kalas2005}
{Kalas}, P. 2005, \apjl, 635, L169

\bibitem[{{Kalas} {et~al.}(2007){Kalas}, {Fitzgerald}, \&
  {Graham}}]{kalas+2007a}
{Kalas}, P., {Fitzgerald}, M.~P., \& {Graham}, J.~R. 2007, \apjl, 661, L85

\bibitem[{{Kalas} {et~al.}(2005){Kalas}, {Graham}, \& {Clampin}}]{kalas+2005}
{Kalas}, P., {Graham}, J.~R., \& {Clampin}, M. 2005, \nat, 435, 1067

\bibitem[{{Kalas} {et~al.}(2013){Kalas}, {Graham}, {Fitzgerald}, \&
  {Clampin}}]{kalas+2013}
{Kalas}, P., {Graham}, J.~R., {Fitzgerald}, M.~P., \& {Clampin}, M. 2013, \apj,
  775, 56

\bibitem[{{Kral} {et~al.}(2015){Kral}, {Th{\'e}bault}, {Augereau},
  {Boccaletti}, \& {Charnoz}}]{kral+2015}
{Kral}, Q., {Th{\'e}bault}, P., {Augereau}, J.-C., {Boccaletti}, A., \&
  {Charnoz}, S. 2015, \aap, 573, A39

\bibitem[{{Kral} {et~al.}(2013){Kral}, {Th{\'e}bault}, \&
  {Charnoz}}]{kral+2013}
{Kral}, Q., {Th{\'e}bault}, P., \& {Charnoz}, S. 2013, \aap, 558, A121

\bibitem[{{Kres{\'a}k}(1976)}]{kresak1976}
{Kres{\'a}k}, L. 1976, Bulletin of the Astronomical Institutes of
  Czechoslovakia, 27, 35

\bibitem[{{Krist} {et~al.}(2012){Krist}, {Stapelfeldt}, {Bryden}, \&
  {Plavchan}}]{krist+2012}
{Krist}, J.~E., {Stapelfeldt}, K.~R., {Bryden}, G., \& {Plavchan}, P. 2012,
  \aj, 144, 45

\bibitem[{{Krivov} {et~al.}(2013){Krivov}, {Eiroa}, {L{\"o}hne}, {Marshall},
  {Montesinos}, {del Burgo}, {Absil}, {Ardila}, {Augereau}, {Bayo}, {Bryden},
  {Danchi}, {Ertel}, {Lebreton}, {Liseau}, {Mora}, {Mustill}, {Mutschke},
  {Neuh{\"a}user}, {Pilbratt}, {Roberge}, {Schmidt}, {Stapelfeldt},
  {Th{\'e}bault}, {Vitense}, {White}, \& {Wolf}}]{krivov+2013}
{Krivov}, A.~V., {Eiroa}, C., {L{\"o}hne}, T., {et~al.} 2013, \apj, 772, 32

\bibitem[{{Krivov} {et~al.}(2006){Krivov}, {L{\"o}hne}, \& {Srem{\v
  c}evi{\'c}}}]{krivov+2006}
{Krivov}, A.~V., {L{\"o}hne}, T., \& {Srem{\v c}evi{\'c}}, M. 2006, \aap, 455,
  509

\bibitem[{{Kuchner} \& {Stark}(2010)}]{kuchner+stark2010}
{Kuchner}, M.~J. \& {Stark}, C.~C. 2010, \aj, 140, 1007

\bibitem[{{Lee} \& {Chiang}(2016)}]{lee+chiang2016}
{Lee}, E.~J. \& {Chiang}, E. 2016, \apj, 827, 125

\bibitem[{{Leinert} {et~al.}(1983){Leinert}, {Roser}, \&
  {Buitrago}}]{leinert+1983}
{Leinert}, C., {Roser}, S., \& {Buitrago}, J. 1983, \aap, 118, 345

\bibitem[{{Levison} {et~al.}(2012){Levison}, {Duncan}, \&
  {Thommes}}]{levison+2012}
{Levison}, H.~F., {Duncan}, M.~J., \& {Thommes}, E. 2012, \aj, 144, 119

\bibitem[{{Li} \& {Greenberg}(1998)}]{li+greenberg1998}
{Li}, A. \& {Greenberg}, J.~M. 1998, \aap, 331, 291

\bibitem[{{Liou} \& {Zook}(1999)}]{liou+zook1999}
{Liou}, J.-C. \& {Zook}, H.~A. 1999, \aj, 118, 580

\bibitem[{{L{\"o}hne} {et~al.}(2012){L{\"o}hne}, {Augereau}, {Ertel},
  {Marshall}, {Eiroa}, {Mora}, {Absil}, {Stapelfeldt}, {Th{\'e}bault}, {Bayo},
  {Del Burgo}, {Danchi}, {Krivov}, {Lebreton}, {Letawe}, {Magain}, {Maldonado},
  {Montesinos}, {Pilbratt}, {White}, \& {Wolf}}]{loehne+2012}
{L{\"o}hne}, T., {Augereau}, J.-C., {Ertel}, S., {et~al.} 2012, \aap, 537, A110

\bibitem[{{L{\"o}hne} {et~al.}(2008){L{\"o}hne}, {Krivov}, \&
  {Rodmann}}]{loehne+2008}
{L{\"o}hne}, T., {Krivov}, A.~V., \& {Rodmann}, J. 2008, \apj, 673, 1123

\bibitem[{{Mamajek}(2012)}]{mamajek2012}
{Mamajek}, E.~E. 2012, \apjl, 754, L20

\bibitem[{{Min} {et~al.}(2010){Min}, {Kama}, {Dominik}, \& {Waters}}]{min+2010}
{Min}, M., {Kama}, M., {Dominik}, C., \& {Waters}, L.~B.~F.~M. 2010, \aap, 509,
  L6+

\bibitem[{{M{\"u}ller} {et~al.}(2010){M{\"u}ller}, {L{\"o}hne}, \&
  {Krivov}}]{mueller+2010}
{M{\"u}ller}, S., {L{\"o}hne}, T., \& {Krivov}, A.~V. 2010, \apj, 708, 1728

\bibitem[{{Murray} \& {Dermott}(2000)}]{murray+dermott2000}
{Murray}, C.~D. \& {Dermott}, S.~F. 2000, {Solar System Dynamics}

\bibitem[{{Mustill} \& {Wyatt}(2009)}]{mustill+wyatt2009}
{Mustill}, A.~J. \& {Wyatt}, M.~C. 2009, \mnras, 399, 1403

\bibitem[{{Nesvold} \& {Kuchner}(2015{\natexlab{a}})}]{nesvold+kuchner2015b}
{Nesvold}, E.~R. \& {Kuchner}, M.~J. 2015{\natexlab{a}}, \apj, 815, 61

\bibitem[{{Nesvold} \& {Kuchner}(2015{\natexlab{b}})}]{nesvold+kuchner2015a}
{Nesvold}, E.~R. \& {Kuchner}, M.~J. 2015{\natexlab{b}}, \apj, 798, 83

\bibitem[{{Nesvold} {et~al.}(2013){Nesvold}, {Kuchner}, {Rein}, \&
  {Pan}}]{nesvold+2013}
{Nesvold}, E.~R., {Kuchner}, M.~J., {Rein}, H., \& {Pan}, M. 2013, \apj, 777,
  144

\bibitem[{{Nesvold} {et~al.}(2017){Nesvold}, {Naoz}, \&
  {Fitzgerald}}]{nesvold+2017}
{Nesvold}, E.~R., {Naoz}, S., \& {Fitzgerald}, M.~P. 2017, \apjl, 837, L6

\bibitem[{{Nesvold} {et~al.}(2016){Nesvold}, {Naoz}, {Vican}, \&
  {Farr}}]{nesvold+2016}
{Nesvold}, E.~R., {Naoz}, S., {Vican}, L., \& {Farr}, W.~M. 2016, \apj, 826, 19

\bibitem[{{O'Brien} \& {Greenberg}(2003)}]{o'brien+greenberg2003}
{O'Brien}, D.~P. \& {Greenberg}, R. 2003, Icarus, 164, 334

\bibitem[{{Olofsson} {et~al.}(2016){Olofsson}, {Samland}, {Avenhaus},
  {Caceres}, {Henning}, {Mo{\'o}r}, {Milli}, {Canovas}, {Quanz}, {Schreiber},
  {Augereau}, {Bayo}, {Bazzon}, {Beuzit}, {Boccaletti}, {Buenzli}, {Casassus},
  {Chauvin}, {Dominik}, {Desidera}, {Feldt}, {Gratton}, {Janson}, {Lagrange},
  {Langlois}, {Lannier}, {Maire}, {Mesa}, {Pinte}, {Rouan}, {Salter},
  {Thalmann}, \& {Vigan}}]{olofsson+2016}
{Olofsson}, J., {Samland}, M., {Avenhaus}, H., {et~al.} 2016, \aap, 591, A108

\bibitem[{{Pan} {et~al.}(2016){Pan}, {Nesvold}, \& {Kuchner}}]{pan+2016}
{Pan}, M., {Nesvold}, E.~R., \& {Kuchner}, M.~J. 2016, \apj, 832, 81

\bibitem[{{Pan} \& {Schlichting}(2012)}]{pan+schlichting2012}
{Pan}, M. \& {Schlichting}, H.~E. 2012, \apj, 747, 113

\bibitem[{{Pollack} \& {Cuzzi}(1980)}]{pollack+cuzzi1980}
{Pollack}, J.~B. \& {Cuzzi}, J.~N. 1980, Journal of Atmospheric Sciences, 37,
  868

\bibitem[{{Reidemeister} {et~al.}(2011){Reidemeister}, {Krivov}, {Stark},
  {Augereau}, {L{\"o}hne}, \& {M{\"u}ller}}]{reidemeister+2011}
{Reidemeister}, M., {Krivov}, A.~V., {Stark}, C.~C., {et~al.} 2011, \aap, 527,
  A57

\bibitem[{{Robertson}(1937)}]{robertson1937}
{Robertson}, H.~P. 1937, \mnras, 97, 423

\bibitem[{{Schneider} {et~al.}(2014){Schneider}, {Grady}, {Hines}, {Stark},
  {Debes}, {Carson}, {Kuchner}, {Perrin}, {Weinberger}, {Wisniewski},
  {Silverstone}, {Jang-Condell}, {Henning}, {Woodgate}, {Serabyn},
  {Moro-Martin}, {Tamura}, {Hinz}, \& {Rodigas}}]{schneider+2014}
{Schneider}, G., {Grady}, C.~A., {Hines}, D.~C., {et~al.} 2014, \aj, 148, 59

\bibitem[{{Schneider} {et~al.}(2006){Schneider}, {Silverstone}, {Hines},
  {Augereau}, {Pinte}, {M{\'e}nard}, {Krist}, {Clampin}, {Grady}, {Golimowski},
  {Ardila}, {Henning}, {Wolf}, \& {Rodmann}}]{schneider+2006}
{Schneider}, G., {Silverstone}, M.~D., {Hines}, D.~C., {et~al.} 2006, \apj,
  650, 414

\bibitem[{{Shannon} {et~al.}(2014){Shannon}, {Clarke}, \&
  {Wyatt}}]{shannon+2014}
{Shannon}, A., {Clarke}, C., \& {Wyatt}, M. 2014, \mnras, 442, 142

\bibitem[{{Stark} \& {Kuchner}(2008)}]{stark+kuchner2008}
{Stark}, C.~C. \& {Kuchner}, M.~J. 2008, \apj, 686, 637

\bibitem[{{Stark} \& {Kuchner}(2009)}]{stark+kuchner2009}
{Stark}, C.~C. \& {Kuchner}, M.~J. 2009, \apj, 707, 543

\bibitem[{{Stewart} \& {Leinhardt}(2009)}]{stewart+leinhardt2009}
{Stewart}, S.~T. \& {Leinhardt}, Z.~M. 2009, \apjl, 691, L133

\bibitem[{Strubbe \& Chiang(2006)}]{strubbe+chiang2006}
Strubbe, L.~E. \& Chiang, E.~I. 2006, \apj, 648, 652

\bibitem[{{Thalmann} {et~al.}(2011){Thalmann}, {Janson}, {Buenzli}, {Brandt},
  {Wisniewski}, {Moro-Mart{\'{\i}}n}, {Usuda}, {Schneider}, {Carson},
  {McElwain}, {Grady}, {Goto}, {Abe}, {Brandner}, {Dominik}, {Egner}, {Feldt},
  {Fukue}, {Golota}, {Guyon}, {Hashimoto}, {Hayano}, {Hayashi}, {Hayashi},
  {Henning}, {Hodapp}, {Ishii}, {Iye}, {Kandori}, {Knapp}, {Kudo}, {Kusakabe},
  {Kuzuhara}, {Matsuo}, {Miyama}, {Morino}, {Nishimura}, {Pyo}, {Serabyn},
  {Suto}, {Suzuki}, {Takahashi}, {Takami}, {Takato}, {Terada}, {Tomono},
  {Turner}, {Watanabe}, {Yamada}, {Takami}, \& {Tamura}}]{thalmann+2011}
{Thalmann}, C., {Janson}, M., {Buenzli}, E., {et~al.} 2011, \apjl, 743, L6

\bibitem[{{Th{\'e}bault}(2012)}]{thebault2012}
{Th{\'e}bault}, P. 2012, \aap, 537, A65

\bibitem[{{Th{\'e}bault} \& {Augereau}(2007)}]{thebault+augereau2007}
{Th{\'e}bault}, P. \& {Augereau}, J.-C. 2007, \aap, 472, 169

\bibitem[{{Th{\'e}bault} {et~al.}(2003){Th{\'e}bault}, {Augereau}, \&
  {Beust}}]{thebault+2003}
{Th{\'e}bault}, P., {Augereau}, J.-C., \& {Beust}, H. 2003, \aap, 408, 775

\bibitem[{{Th{\'e}bault} {et~al.}(2014){Th{\'e}bault}, {Kral}, \&
  {Augereau}}]{thebault+2014}
{Th{\'e}bault}, P., {Kral}, Q., \& {Augereau}, J.-C. 2014, \aap, 561, A16

\bibitem[{{Th{\'e}bault} {et~al.}(2012){Th{\'e}bault}, {Kral}, \&
  {Ertel}}]{thebault+2012}
{Th{\'e}bault}, P., {Kral}, Q., \& {Ertel}, S. 2012, \aap, 547, A92

\bibitem[{{Th{\'e}bault} {et~al.}(2010){Th{\'e}bault}, {Marzari}, \&
  {Augereau}}]{thebault+2010}
{Th{\'e}bault}, P., {Marzari}, F., \& {Augereau}, J.-C. 2010, \aap, 524, A13

\bibitem[{{Th{\'e}bault} \& {Wu}(2008)}]{thebault+wu2008}
{Th{\'e}bault}, P. \& {Wu}, Y. 2008, \aap, 481, 713

\bibitem[{{Thilliez} \& {Maddison}(2015)}]{thilliez+maddison2015}
{Thilliez}, E. \& {Maddison}, S.~T. 2015, \pasa, 32, e039

\bibitem[{{Vitense} {et~al.}(2012){Vitense}, {Krivov}, {Kobayashi}, \&
  {L{\"o}hne}}]{vitense+2012}
{Vitense}, C., {Krivov}, A.~V., {Kobayashi}, H., \& {L{\"o}hne}, T. 2012, \aap,
  540, A30

\bibitem[{{Vitense} {et~al.}(2014){Vitense}, {Krivov}, \&
  {L{\"o}hne}}]{vitense+2014}
{Vitense}, C., {Krivov}, A.~V., \& {L{\"o}hne}, T. 2014, \aj, 147, 154

\bibitem[{{Wyatt}(2005)}]{wyatt2005a}
{Wyatt}, M.~C. 2005, \aap, 433, 1007

\bibitem[{{Wyatt} {et~al.}(2011){Wyatt}, {Clarke}, \& {Booth}}]{wyatt+2011}
{Wyatt}, M.~C., {Clarke}, C.~J., \& {Booth}, M. 2011, Celestial Mechanics and
  Dynamical Astronomy, 111, 1

\bibitem[{{Wyatt} {et~al.}(1999){Wyatt}, {Dermott}, {Telesco}, {Fisher},
  {Grogan}, {Holmes}, \& {Pi{\~n}a}}]{wyatt+1999}
{Wyatt}, M.~C., {Dermott}, S.~F., {Telesco}, C.~M., {et~al.} 1999, \apj, 527,
  918

\bibitem[{{Wyatt} \& {Whipple}(1950)}]{wyatt+whipple1950}
{Wyatt}, S.~P. \& {Whipple}, F.~L. 1950, \apj, 111, 134

\end{thebibliography}
\end{document}